\documentclass[10pt]{article}
\usepackage{graphicx}
\usepackage{float}
\usepackage{wrapfig}\usepackage{sidecap}
\usepackage[margin=2cm]{caption}
\usepackage{subcaption}
\def\Title#1{\begin{center} {\Large #1 } \end{center}}
\def\Author#1{\begin{center}{ \sc #1} \end{center}}
\def\Address#1{\begin{center}{ \it #1} \end{center}}

\newcommand\pubblock{\rightline{\begin{tabular}{l} Proceedings of the Second Annual LHCP\\ \pubnumber\\
         \pubdate  \end{tabular}}}

\newenvironment{Abstract}{\begin{quotation} \begin{center} 
             \large ABSTRACT \end{center}\bigskip 
      \begin{center}\begin{large}}{\end{large}\end{center} \end{quotation}}

\newenvironment{Presented}{\begin{quotation} \begin{center} 
             PRESENTED AT\end{center}\bigskip 
      \begin{center}\begin{large}}{\end{large}\end{center} \end{quotation}}

\def\beq{\begin{equation}}
\def\eeq#1{\label{#1}\end{equation}}
\def\eeqn{\end{equation}}

\def\beqa{\begin{eqnarray}}
\def\eeqa#1{\label{#1}\end{eqnarray}}
\def\eeqan{\end{eqnarray}}

\let\bar=\overbar

\def\Dslash{\not{\hbox{\kern-4pt $D$}}}
\def\dslash{\not{\hbox{\kern-2pt $\del$}}}

\def\msb{{\bar{\ssstyle M \kern -1pt S}}}

\newcommand{\piz}{{\mathbf{\pi^{0}}}}

\newcommand{\pp}{{pp\,}}

\newcommand{\pbpb}{Pb-Pb\,}

\renewcommand{\aa}{{\rm AA}}

\newcommand{\pt}{{p_{\rm T}}}
\newcommand{\sq}{{\sqrt{s}=}}
\newcommand{\raa}{{R_\aa}}

\newlength{\currentparskip}
\newcommand{\minibox}[2]{%
\setlength{\currentparskip}{\parskip}%
\parbox{#1}{\setlength{\parskip}{\currentparskip}%
#2}%
}

\usepackage{color}

\newcommand\pubnumber{ }
\newcommand\pubdate{\today}

\def\affiliation{
On behalf of the ALICE collaboration, \\
Laboratoire de Physique Subatomique et des Technologies Associ\'{e}es SUBATECH.
\\
4 Rue Alfred Kastler, 44300 Nantes, France}

\begin{document}

% large size for the first page
\large
\begin{titlepage}
\pubblock

%% Change the title, name, abstract
%% Title 
\vfill
\Title{Probing hot and dense matter production in heavy ion collisions via neutral mesons and photons with the ALICE detector at the LHC. }
\vfill

%  if you need to add the support use this, fill the \support definition above. 
%   \Author{ FIRSTNAME LASTNAME \support }
\Author{ Astrid Morreale  }
\Address{\affiliation}
\vfill
\begin{Abstract}

One of the key signatures of the Quark Gluon Plasma (QGP) is the modification of hadron and direct photon spectra in heavy-ion collisions as compared to proton-proton (pp) collisions. Suppression of hadron production at high transverse momenta in heavy-ion collisions can be explained by the energy loss of the partons produced in the hard scattering processes which traverse the hot and dense QCD matter. The dependence of the observed suppression on the transverse momentum ($\pt$) of the measured hadron towards higher $\pt$ is an important input for the theoretical understanding of jet quenching effects in the QGP and the nature of energy loss. Another key observable which has helped establish the energy loss picture, is high $\pt$ direct photon production for which no suppression is expected. For low $\pt$ photon production, it is expected that thermal sources will lead to enhancement of direct photons.

We report an overview of photon and neutral meson production measurements by the ALICE experiment at the LHC in heavy-ion and pp collisions.

\end{Abstract}
\vfill

% DO NOT CHANGE 
\begin{Presented}
The Second Annual Conference\\
 on Large Hadron Collider Physics \\
Columbia University, New York, U.S.A \\ 
June 2-7, 2014
\end{Presented}
\vfill
\end{titlepage}
\def\thefootnote{\fnsymbol{footnote}}
\setcounter{footnote}{0}
\clearpage
% normal size for the rest
\normalsize 

%% Your paper should be entered below. 

\section{Introduction}
Quantum chromodynamics (QCD), the established theory of strongly interacting matter, predicts that above a critical energy density, hadrons, the constituents of nuclear matter at low temperature, decompose into a plasma of quarks and gluons. Such a state is believed to have existed a few microseconds after the Big Bang~\cite{QCD}.
Colliding heavy ions at relativistic energies is a way to study and characterize the QGP. 
ALICE is the only experiment at the LHC optimized to study the QGP~\cite{emcal}.
 
The observed suppression of hadron production at high transverse momenta in heavy-ion collisions 
has been interpreted as energy loss of the scattered parton in the hot and dense strongly-interacting QGP formed in the collision. Photons and mesons are interesting probes of the QGP as their birth typically occurs at different stages of the collision~\cite{QCD2}. In pp collisions, photons are produced at the initial state of collisions while hadrons are produced from parton fragmentation in the QCD vacuum. Within the QGP and in central heavy ion collisions, scattered partons  interact strongly with the medium leading to modifications of parton fragmentation while photons typically do not interact with the medium. Both of these latter effects can be observed via inclusive spectra modifications and hadron-hadron correlations~\cite{QCD2}.
Direct photons are produced not only at the initial stage of the collisions such as in \pp collisions but also during the whole evolution of the QGP in final state scattering processes. Direct photon production at high $\pt$ probes binary $NN$ scaling while at low $\pt$ an excess of direct photons would indicate thermal radiation of the quark gluon plasma.~\cite{thermalphoton}
\vspace{-0.2cm} 
\section{ALICE detection of photons}
The ALICE detector measures photons with three fully complementary methods: photon conversion method (PCM) and electromagnetic shower calorimetry using two detectors: PHOS and EMCal.
The PCM method measures photons and meson yields by reconstructing $e^{+}e^{-}$ pairs proceeding from photon conversions in the material of ALICE inner detectors. The electromagnetic calorimeters PHOS and EMCal are based on energy measurements via total absorption of particles.

PCM measurements use the ALICE inner tracking system (ITS) and the time projection chamber (TPC).
The conversion point can be reconstructed with a $z$ and $\phi$ resolution of 1.5~cm and 7~mrad respectively. 
While photon conversion probability $X/X_{0} =(11.4\pm 0.5_{sys}~\%)$ is  small it is compensated by its wide acceptance: full azimuthal acceptance and 1.8 units of pseudo rapidity coverage.

The PHOS detector~\cite{phos} is a high granularity calorimeter  ($\sigma_{E(\rm GeV)}/E=0.018/E\oplus0.033/\sqrt{E}\oplus0.011$) composed of PbWO$_4$ crystals with 3 modules at 4.6 m from ALICE's interaction point (IP). It covers a pseudorapidity of 0.26 units and has an azimuthal acceptance of $60^\circ$.  

The EMCal detector~\cite{emcal} is a modular sampling calorimeter, composed of 77 alternating layers of 1.4 mm lead  and  1.7 mm scintillator. The EMCal like PHOS is high granularity detector ($\sigma_{E(\rm GeV)}/{E}=0.05/E\oplus 0.1\pm0.04/\sqrt{E}\oplus0.017$) composed of 11 super modules located 4.4 m from ALICE's IP.  It covers a pseudorapidity of 1.4 units and has $100^\circ$ of azimuthal coverage. 

\vspace{-0.1cm}
\hspace{-1cm}
\begin{tabular}{lr}
\minibox{10cm}{
\normalsize
\vspace{-1.0cm}
The three detection methods described above provide independent measurements with completely different systematic uncertainties. Their combined measurements of photon observables are thus important to minimize biases. Fig.~\ref{fig:figure1}  shows the layout of the ALICE detectors including those which are relevant for neutral meson and photon analyses: PHOS, EMCAL, ITS, and TPC.}
&
\minibox{6cm}{
\centering
\captionsetup{width=1.0\linewidth}
\includegraphics[width=6cm]{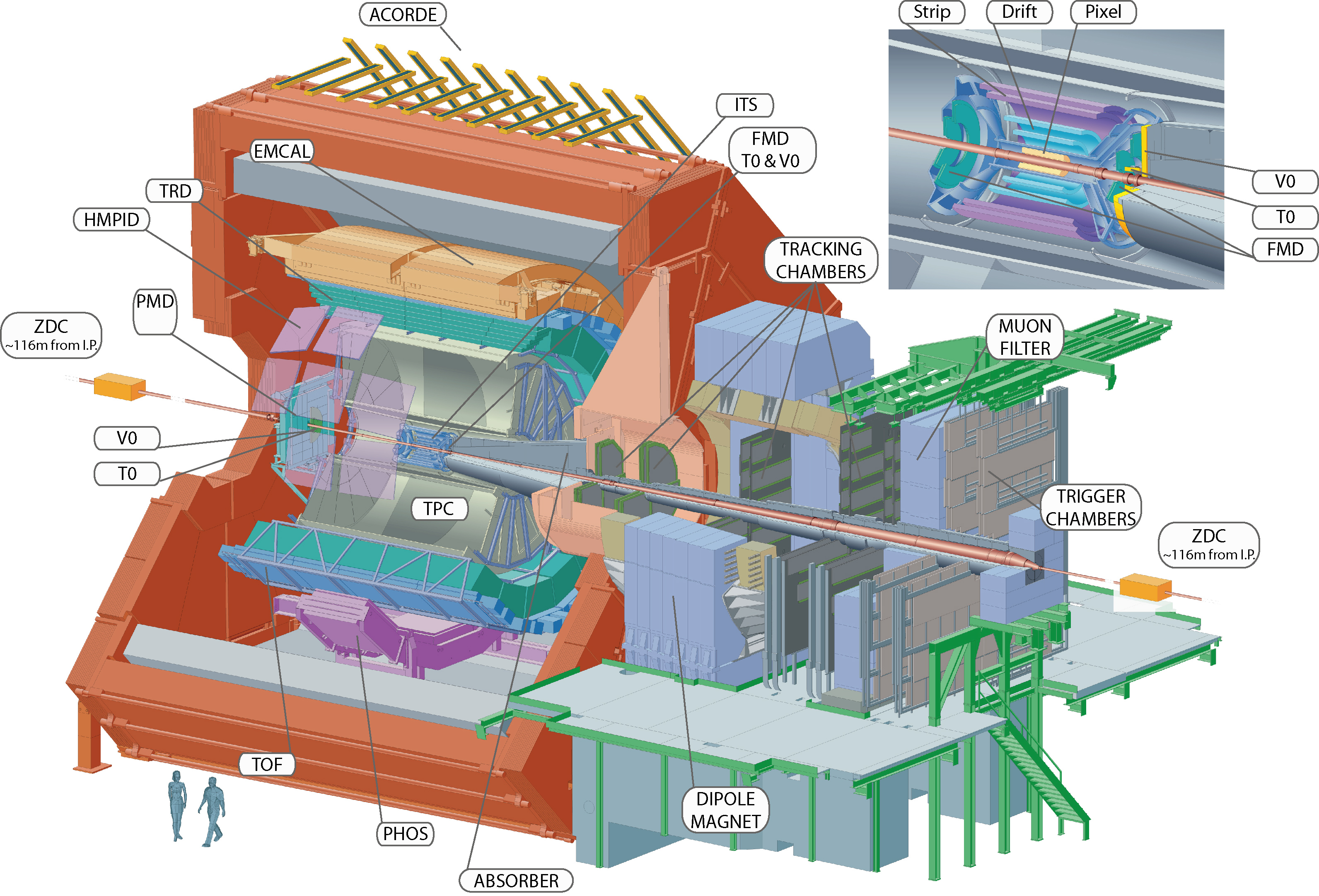}
\captionof{figure}{Layout of ALICE.}
\label{fig:figure1}
}
\end{tabular}

\vspace{-1.2cm}
\section{Neutral meson reconstruction}
Measurement of $\piz$ and $\eta$ mesons is performed by reconstruction the invariant mass $M_{\gamma\gamma}$  in bins of the di-photon pair's $\pt$. Reconstruction of the invariant mass via the PCM method permits meson measurements starting at $\pt>0.4$ GeV/c, while the PHOS and EMCal detectors can provide measurements in the intermediate (above 1~GeV/c) and high $\pt$ (above 10 GeV/c) regions respectively. Fig.~\ref{fig:figure2} exemplifies the invariant mass distributions at different $\pt$ bins illustrating typical $\pt$ values of the the relevant ALICE detectors~\cite{pppublished}. 

\vspace{-0.1cm}
\begin{figure}
\centering
\begin{subfigure}[b]{0.28\textwidth}
\includegraphics[width=4.0cm]{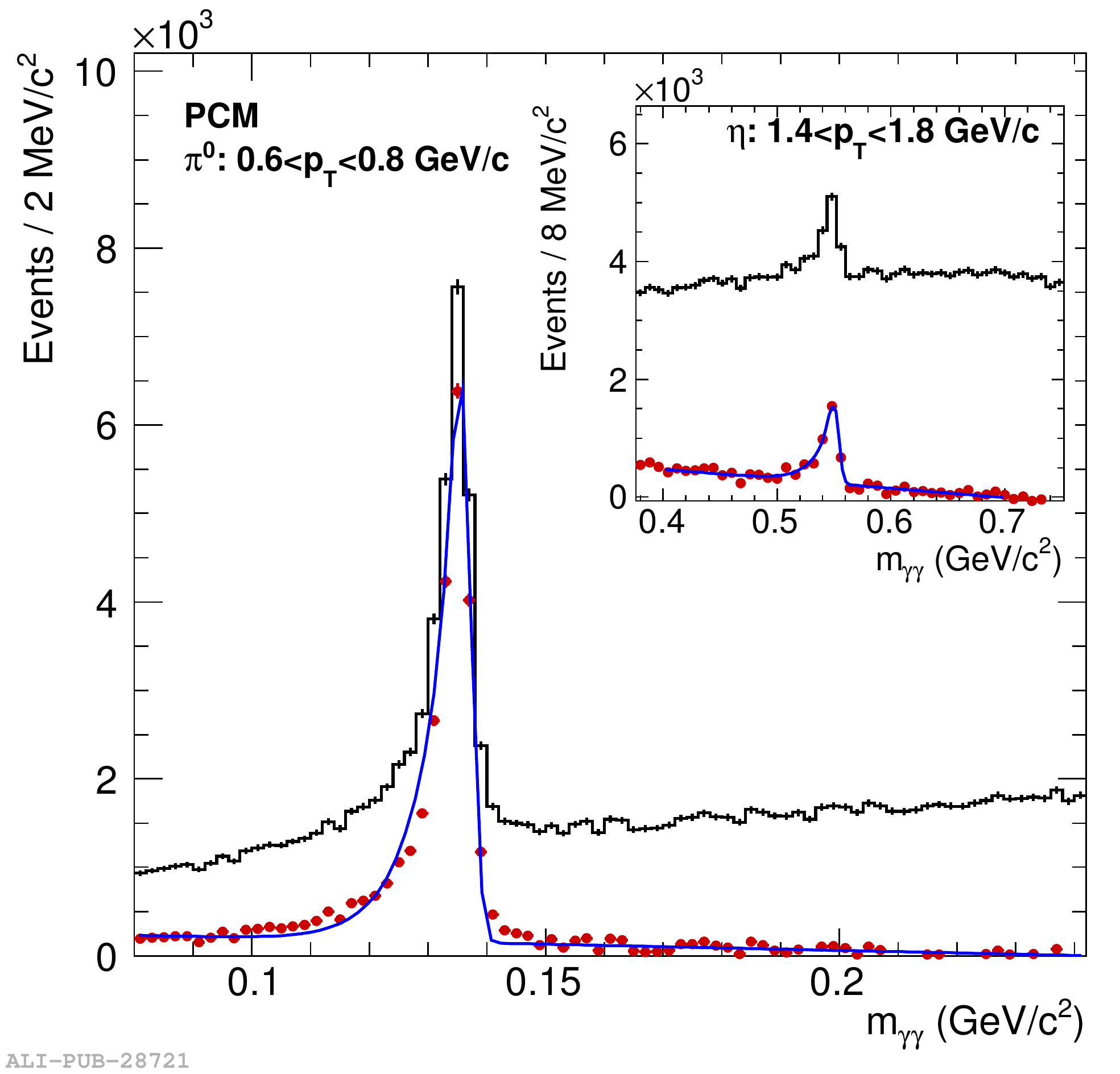}
%\caption{\footnotesize $\piz$ PCM ($\pp$)}\label{fig:pcm1pi0}
\end{subfigure}\begin{subfigure}[b]{0.28\textwidth}
\hspace{-0.6cm}\includegraphics[width=4.0cm]{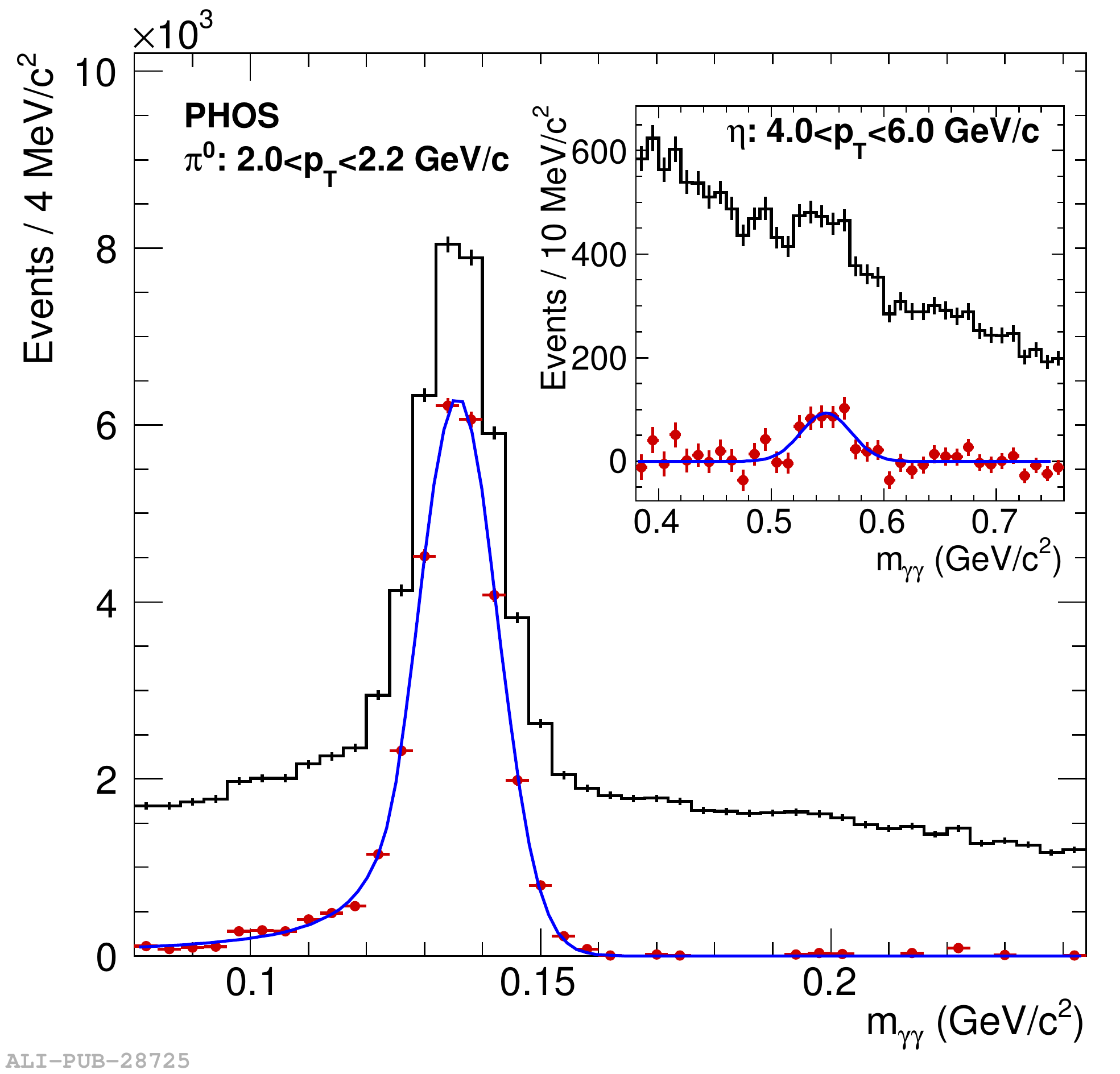}
%\caption{\footnotesize $\piz$ PHOS ($\pp$)}\label{fig:phos1}
\end{subfigure}\begin{subfigure}[b]{0.28\textwidth}
\hspace{-1.2cm}\includegraphics[width=3.9cm]{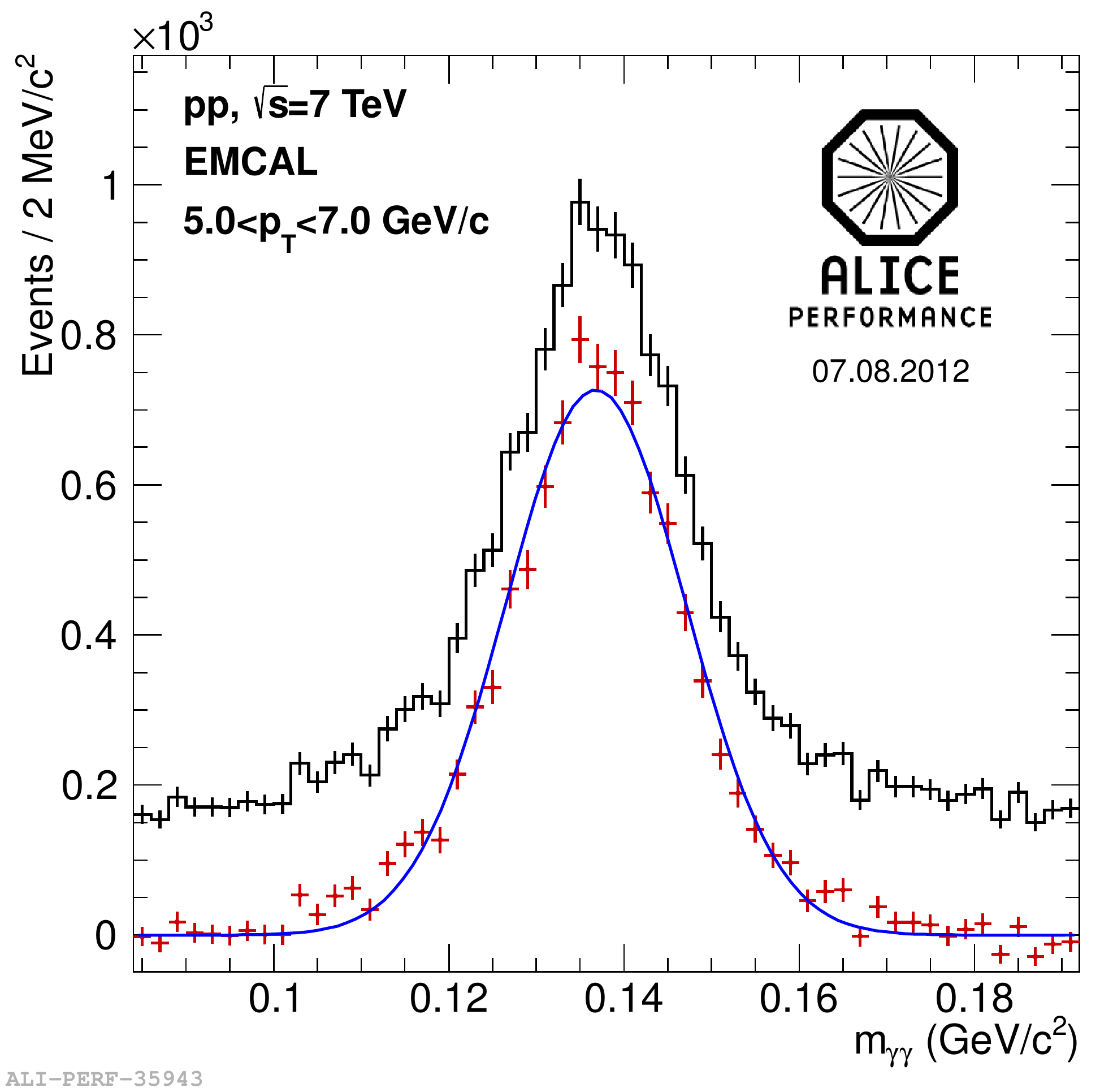}
%\caption{\footnotesize $\piz$ EMCal ($\pp$)}\label{fig:emcalpi0}
\end{subfigure}\begin{subfigure}[b]{0.28\textwidth}
\hspace{-1.2cm}\includegraphics[width=4.0cm]{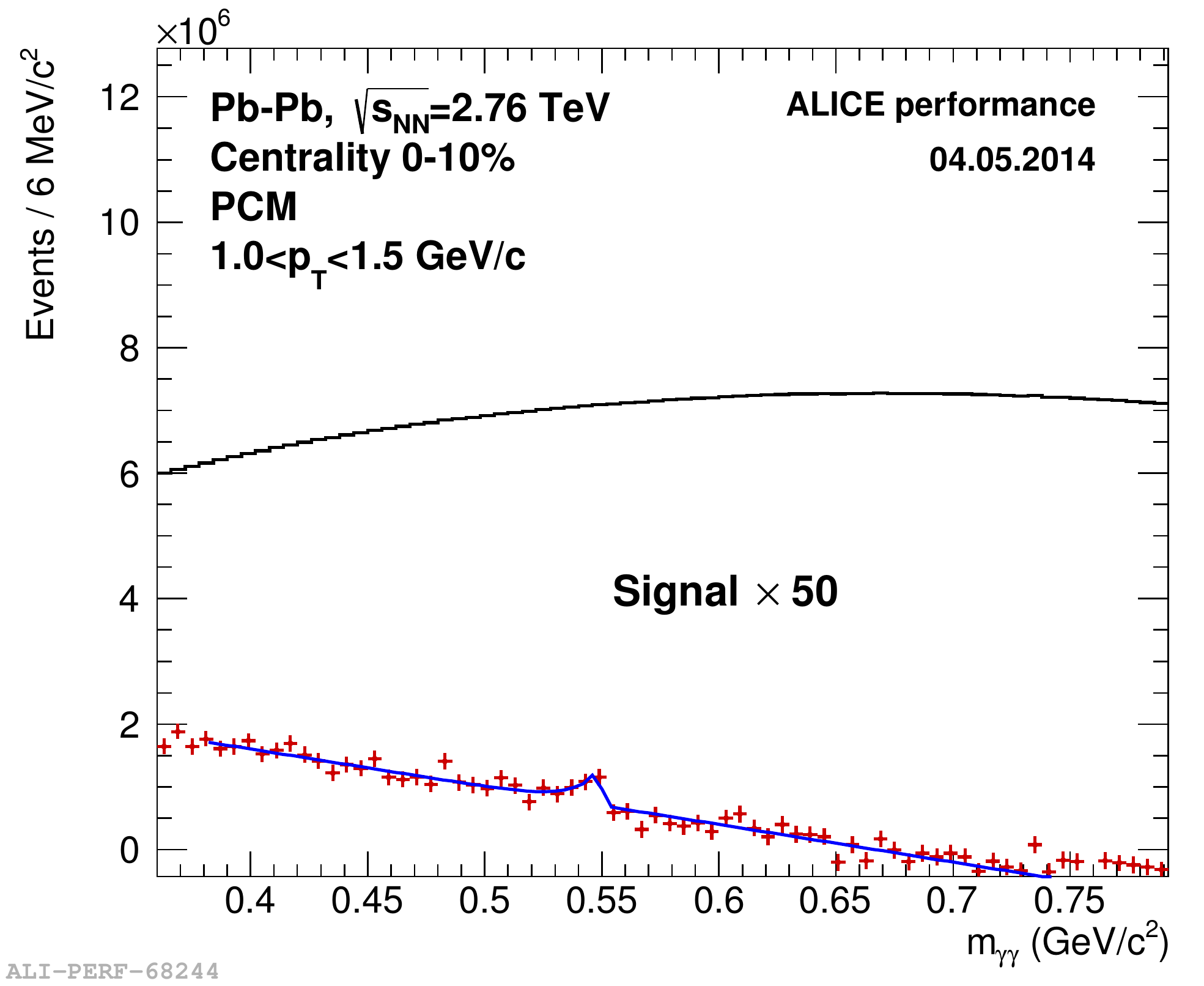}
%\caption{\footnotesize $\eta$ PCM ($\pbpb$)}\label{fig:pcm2eta}
\end{subfigure}

\begin{subfigure}[b]{0.28\textwidth}
\includegraphics[width=4.0cm]{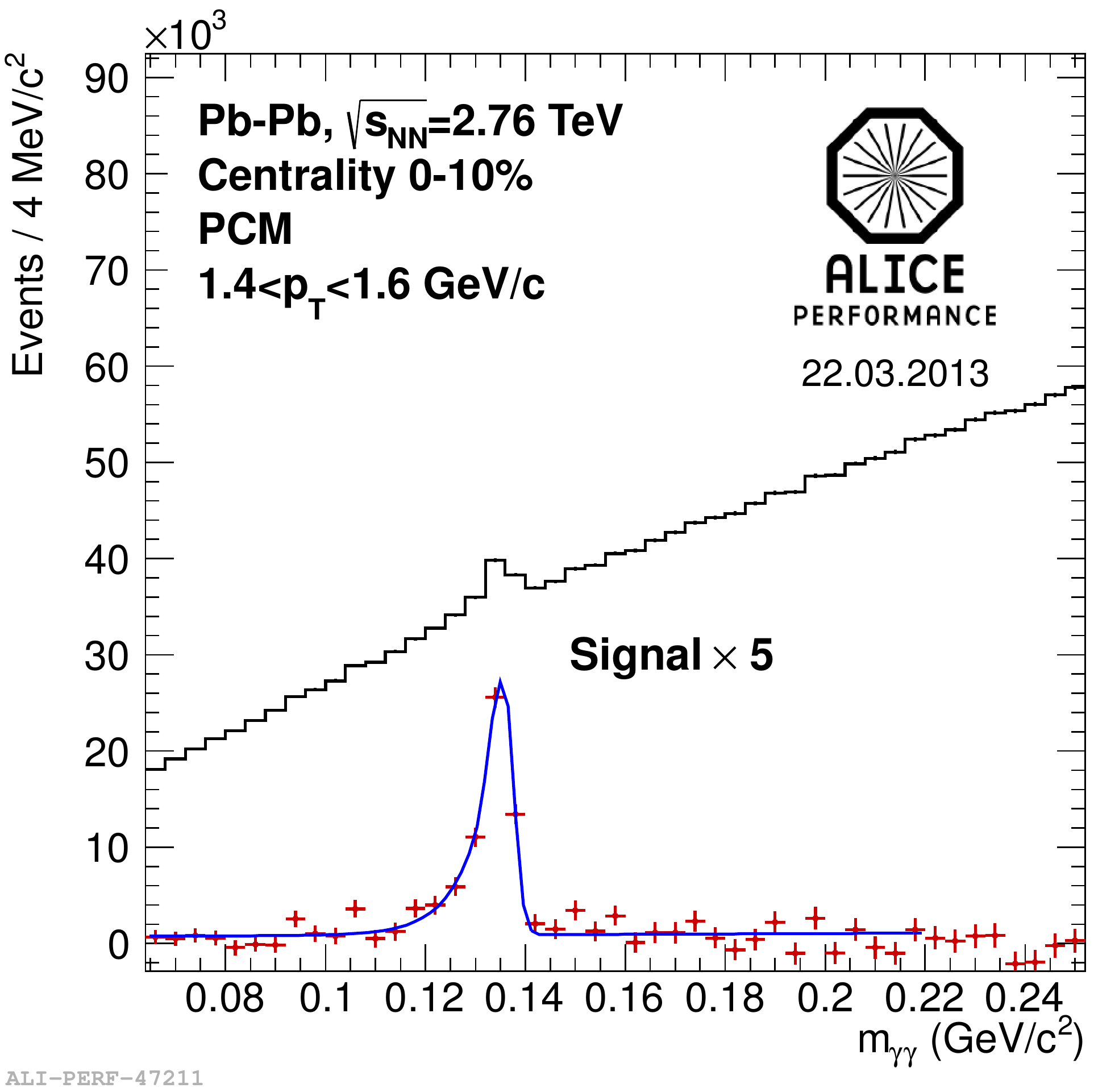}
%\caption{\footnotesize $\piz$ PCM ($\pbpb$)}\label{fig:pcm2pi0}
\end{subfigure}\begin{subfigure}[b]{0.28\textwidth}
\hspace{-0.6cm}\includegraphics[width=4.0cm]{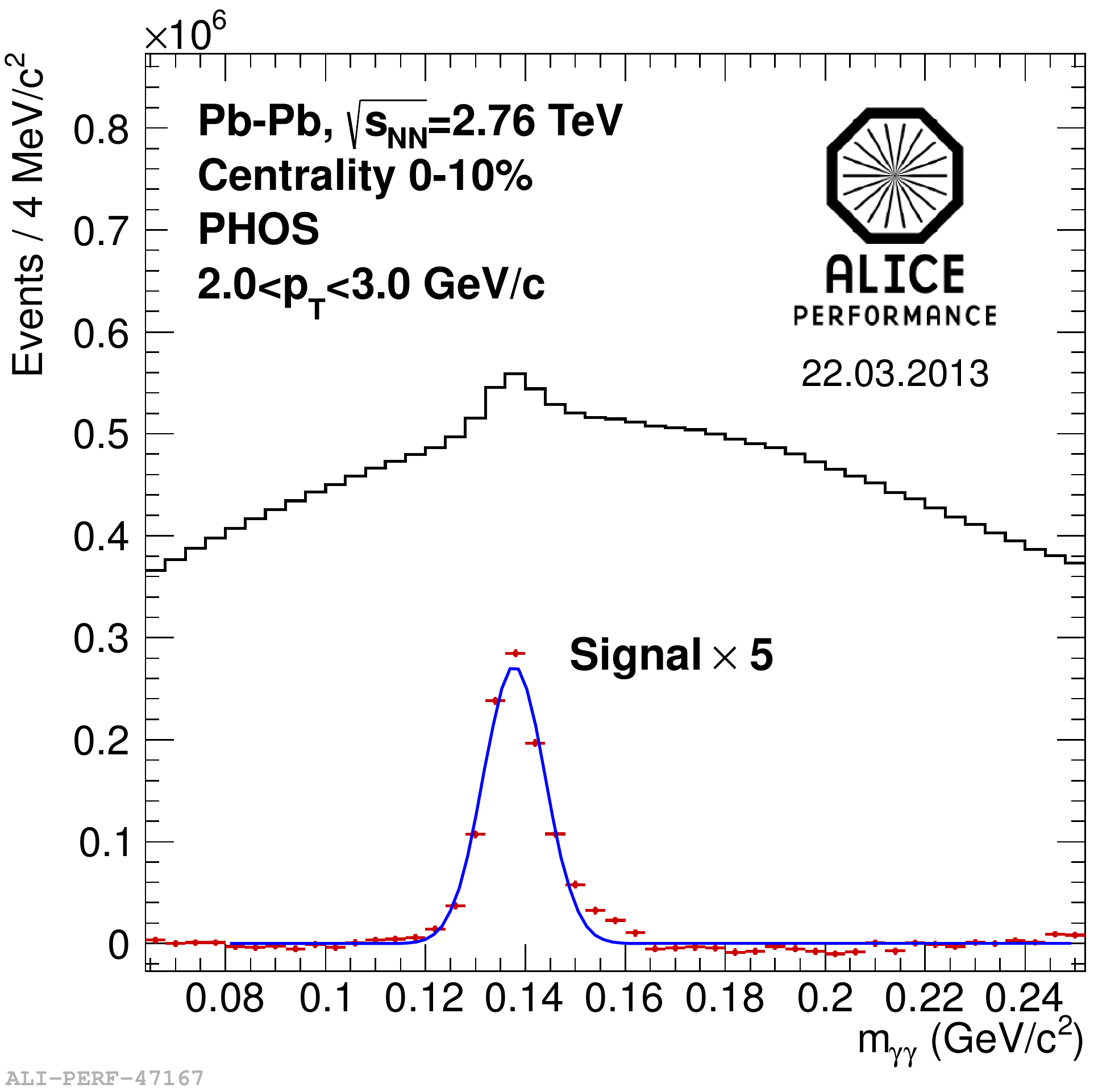}
%\caption{\footnotesize $\piz$ PHOS ($\pbpb$)}\label{fig:phos2}
\end{subfigure}\begin{subfigure}[b]{0.28\textwidth}
\hspace{-1.2cm}\includegraphics[width=4.0cm]{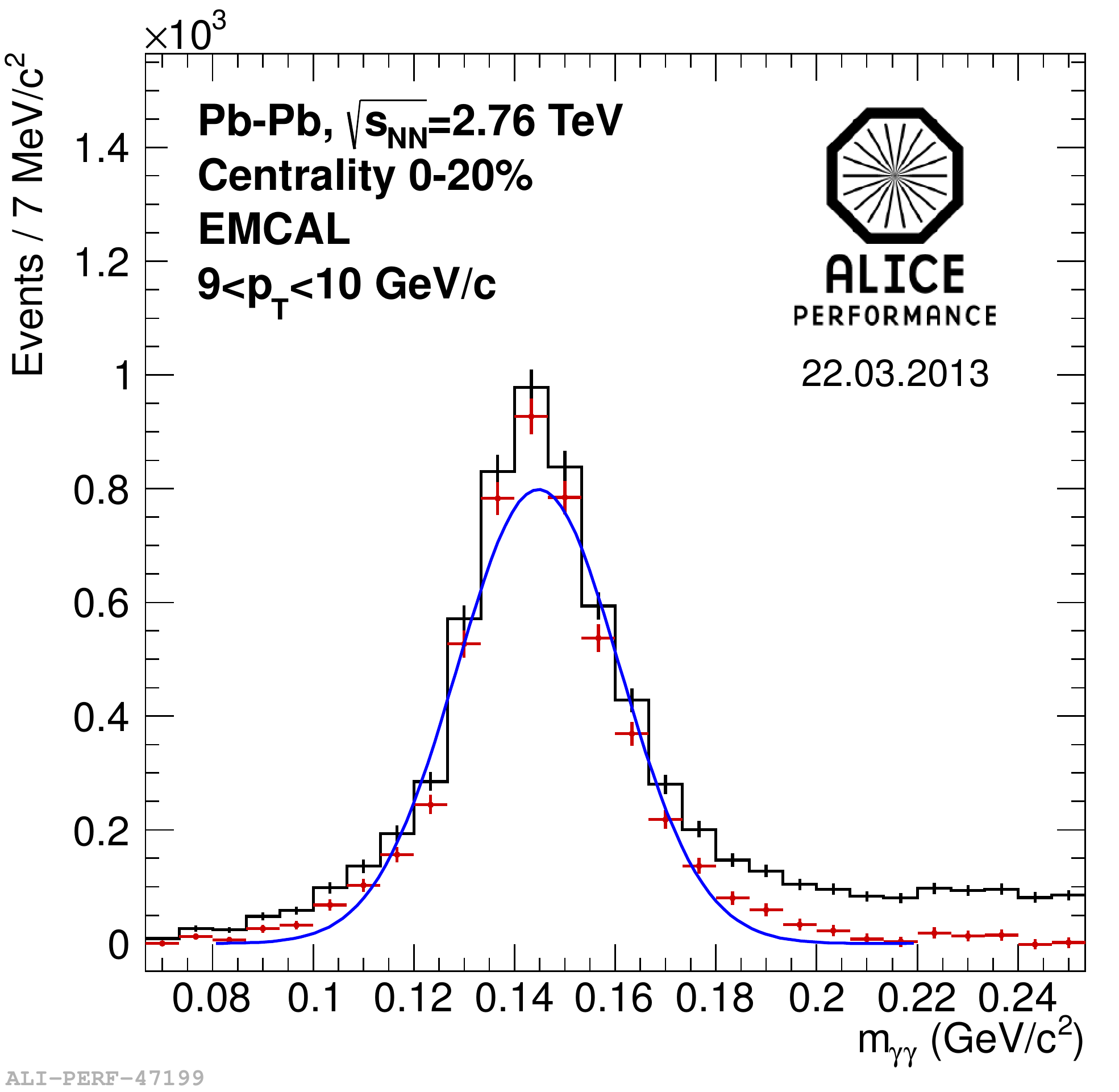}
%\caption{\footnotesize $\piz$ EMCal ($\pbpb$)}\label{fig:emcalpi0}
\end{subfigure}\begin{subfigure}[b]{0.28\textwidth}
\hspace{-1.2cm}\includegraphics[width=4.0cm]{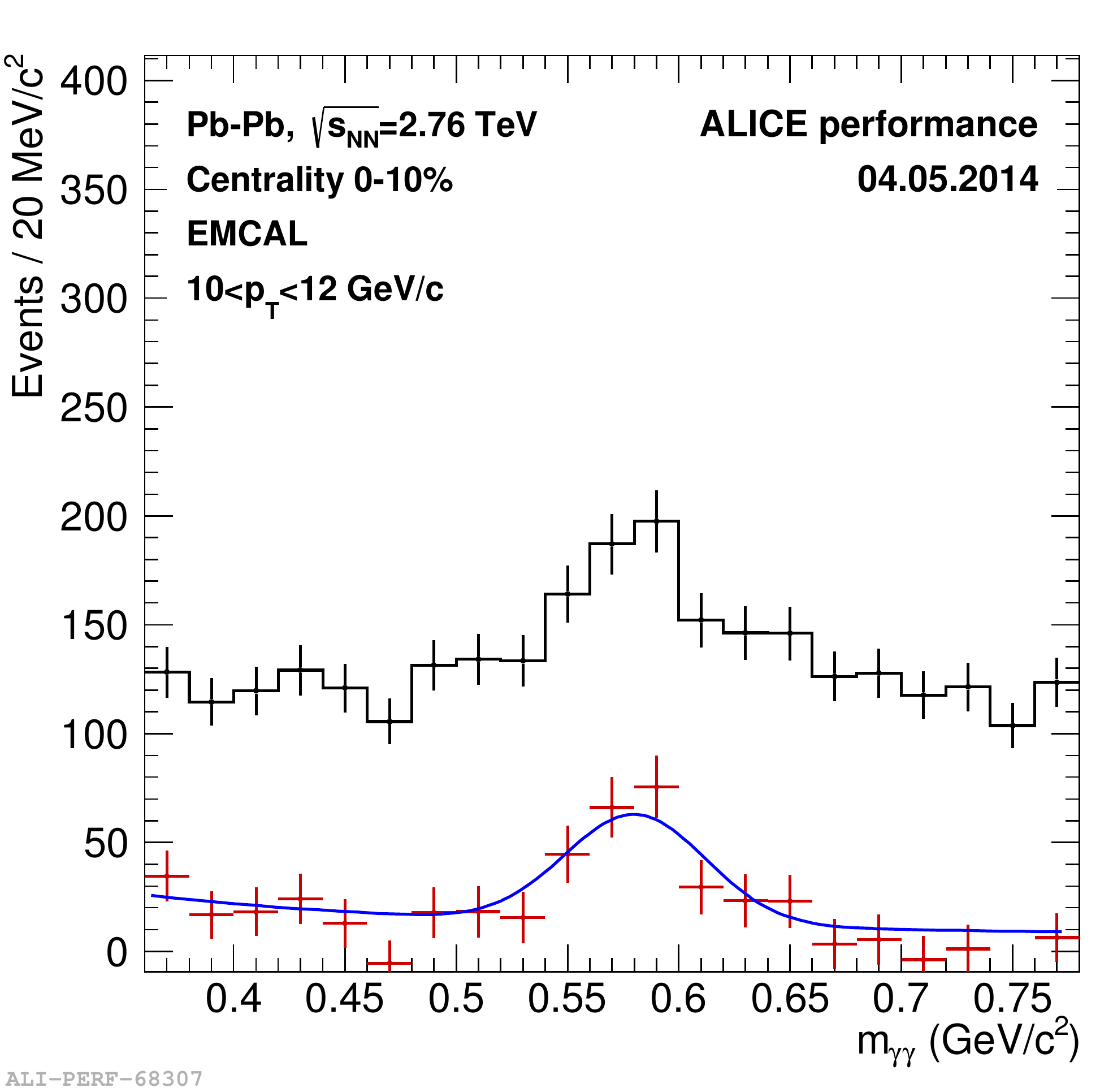}
%\caption{\footnotesize $\eta$ EMCal ($\pbpb$)}\label{fig:emcaleta}
\end{subfigure}\captionsetup{width=1.0\linewidth}\vspace{-0.4cm}
\captionof{figure}{\footnotesize  Invariant masses of neutral mesons measured by ALICE in pp and $\pbpb$ collisions~\cite{pppublished}.}
\end{figure}\label{fig:figure2}
\vspace{-0.5cm}
%%%%%%%%%%%%%
\vspace{0.3cm}
\section{$\piz$ invariant yields in pp at $\sqrt{s} =$ 0.9, 2.76 and 7 TeV}
\begin{figure}[H]
 \centering
 \begin{subfigure}[b]{0.4\textwidth}
 \includegraphics[width=5.2cm]{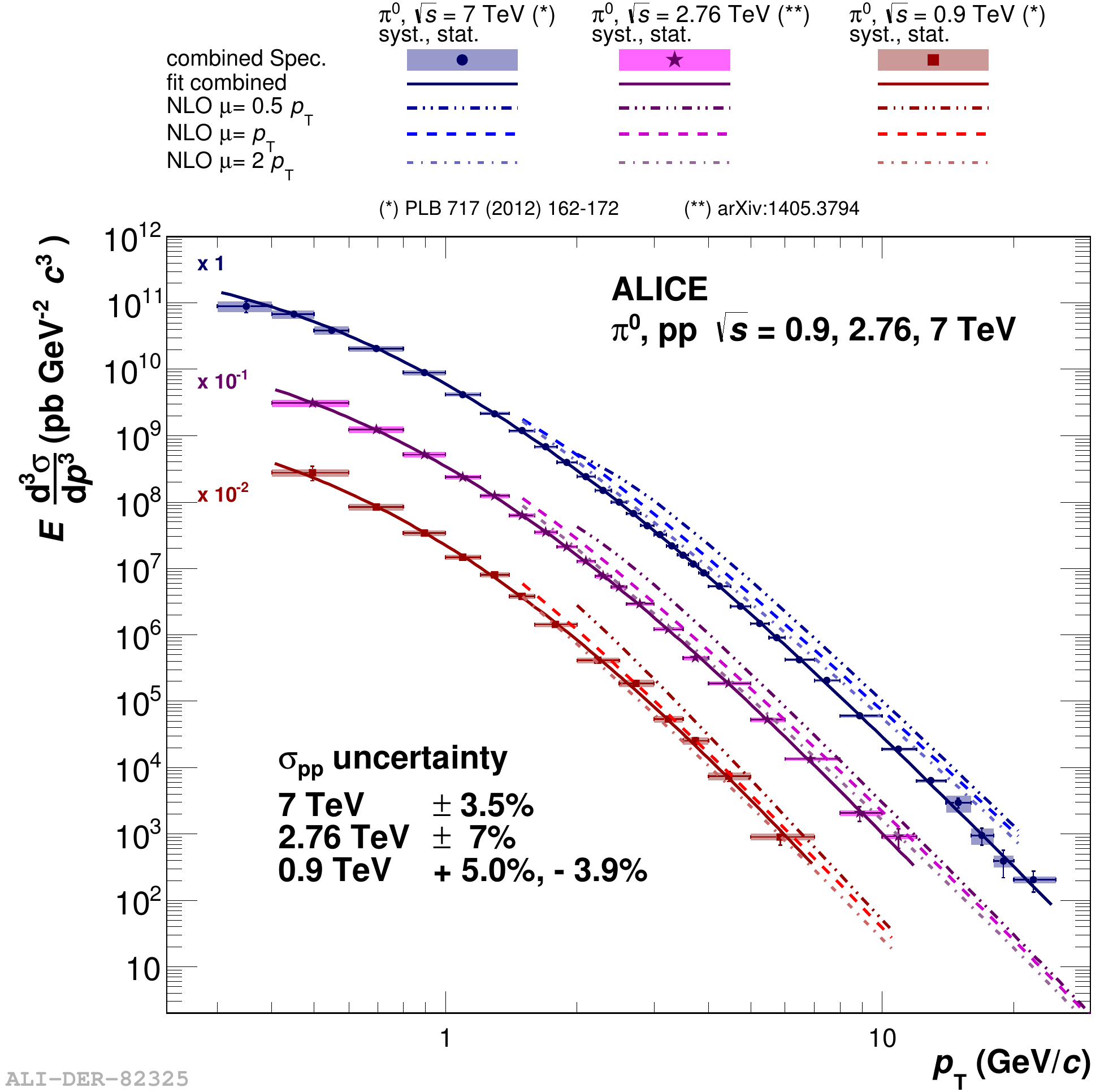}
 \end{subfigure}\begin{subfigure}[b]{0.4\textwidth}
 \hspace{-1mm}\includegraphics[width=5.2cm]{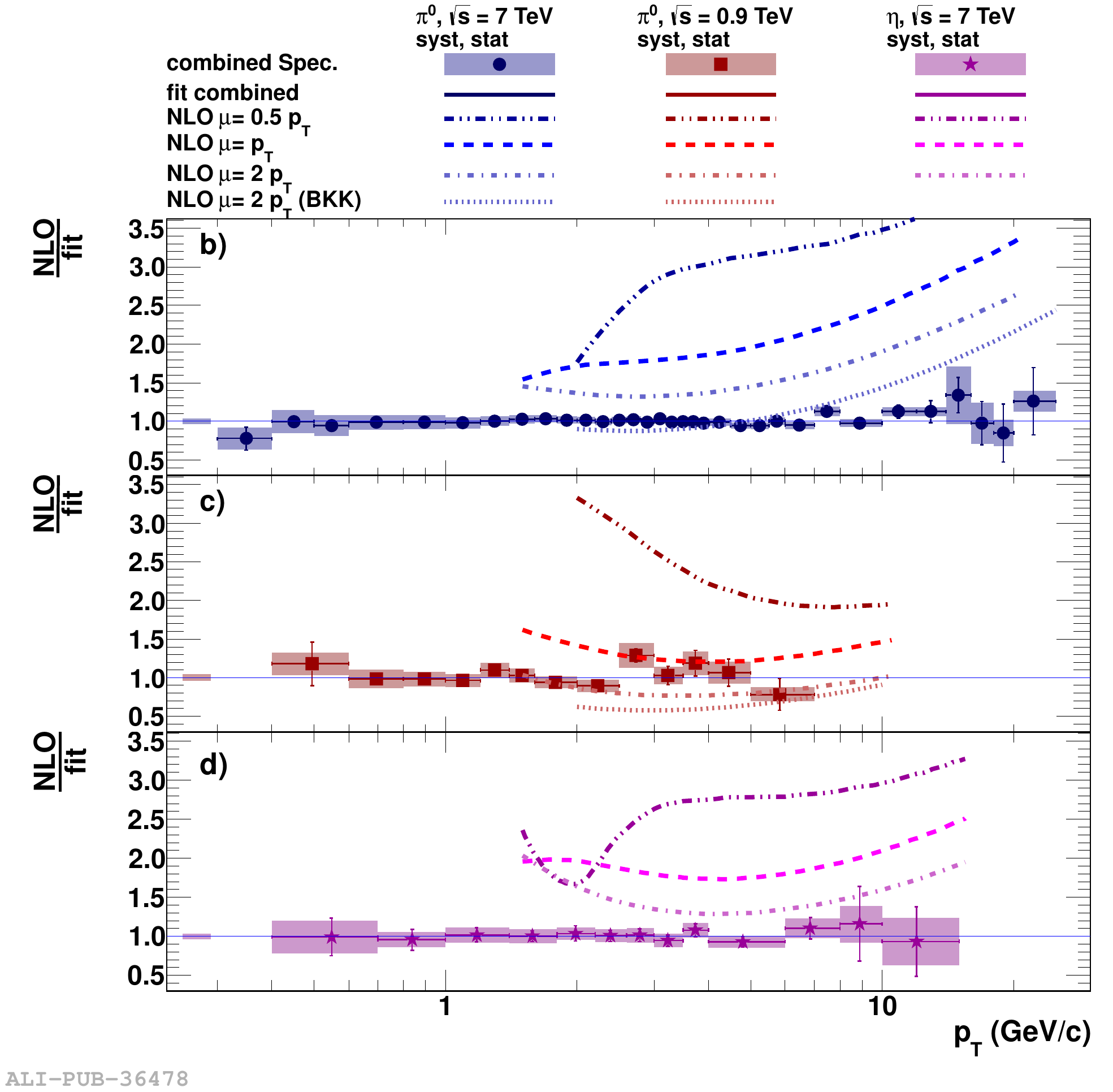}
 \end{subfigure}\vspace{-0.4cm}
\caption{\footnotesize Left: invariant yields at three ${\sqrt{s}}$ compared to pQCD NLO calculations at three different (and equal) factorization, renormalization scales. Right panel: ratio of the three $\piz$ ALICE measurements to pQCD calculations.} \label{fig:ppyields}
%\hspace{1.0cm}
\end{figure}

ALICE has measured $\piz$ invariant yields in pp collisions at three different center of mass energies $\sq$ 0.9, 2.76 and 7 TeV~\cite{pppublished,pbpbpublished} (Fig.~\ref{fig:ppyields}). The measurements presented in Fig.~\ref{fig:ppyields} indicate a power law dependence at high $\pt$ with  a measured power value of $n=6.0\pm0.1$ at $\sq$2.76~TeV, a value lower than what has been observed with measurements performed at lower $\sqrt{s}$~\cite{PHENIX0}. The measurements have additionally been compared to recent next-to-leading order pQCD (NLO pQCD) which have included in their parametrizations CTEQ6M5 parton distribution functions (PDF), DSS and BKK($\piz$)  as well as AESS ($\piz$) fragmentation functions (FF).  These comparisons as well as the ratio are illustrated in Fig.~\ref{fig:ppyields}. A discrepancy of NLO pQCD and the ALICE data at $\sq$2.76~TeV and 7~TeV is observed from these comparisons which may hint towards the FF not yet evolved at this energy. 
\vspace{-0.2cm}

\begin{figure}[H]
\centering
\begin{subfigure}[b]{0.4\textwidth}
\includegraphics[width=5.0cm]{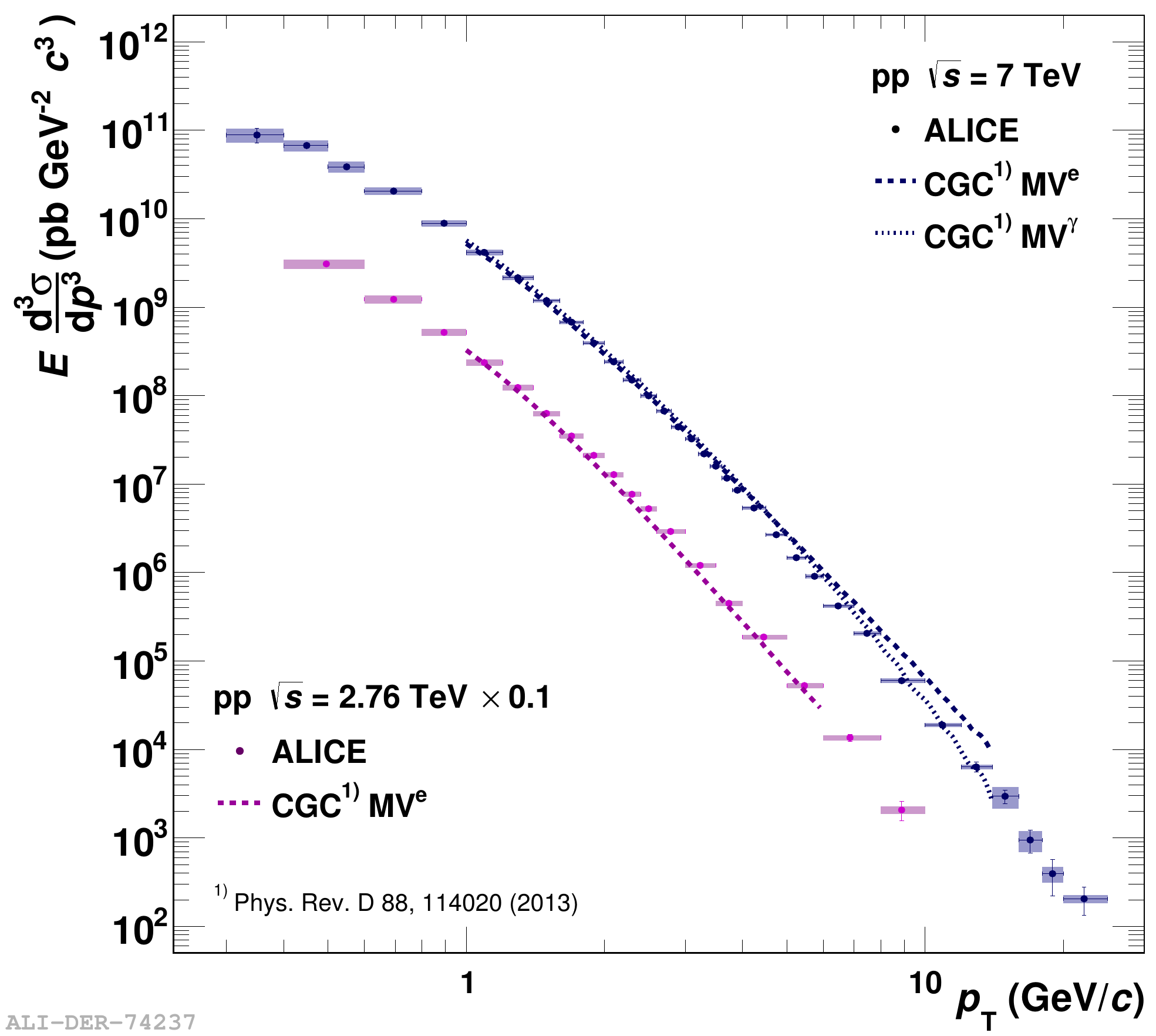}
\end{subfigure}
\begin{subfigure}[b]{0.4\textwidth}
\hspace{-1mm}\includegraphics[width=4.9cm]{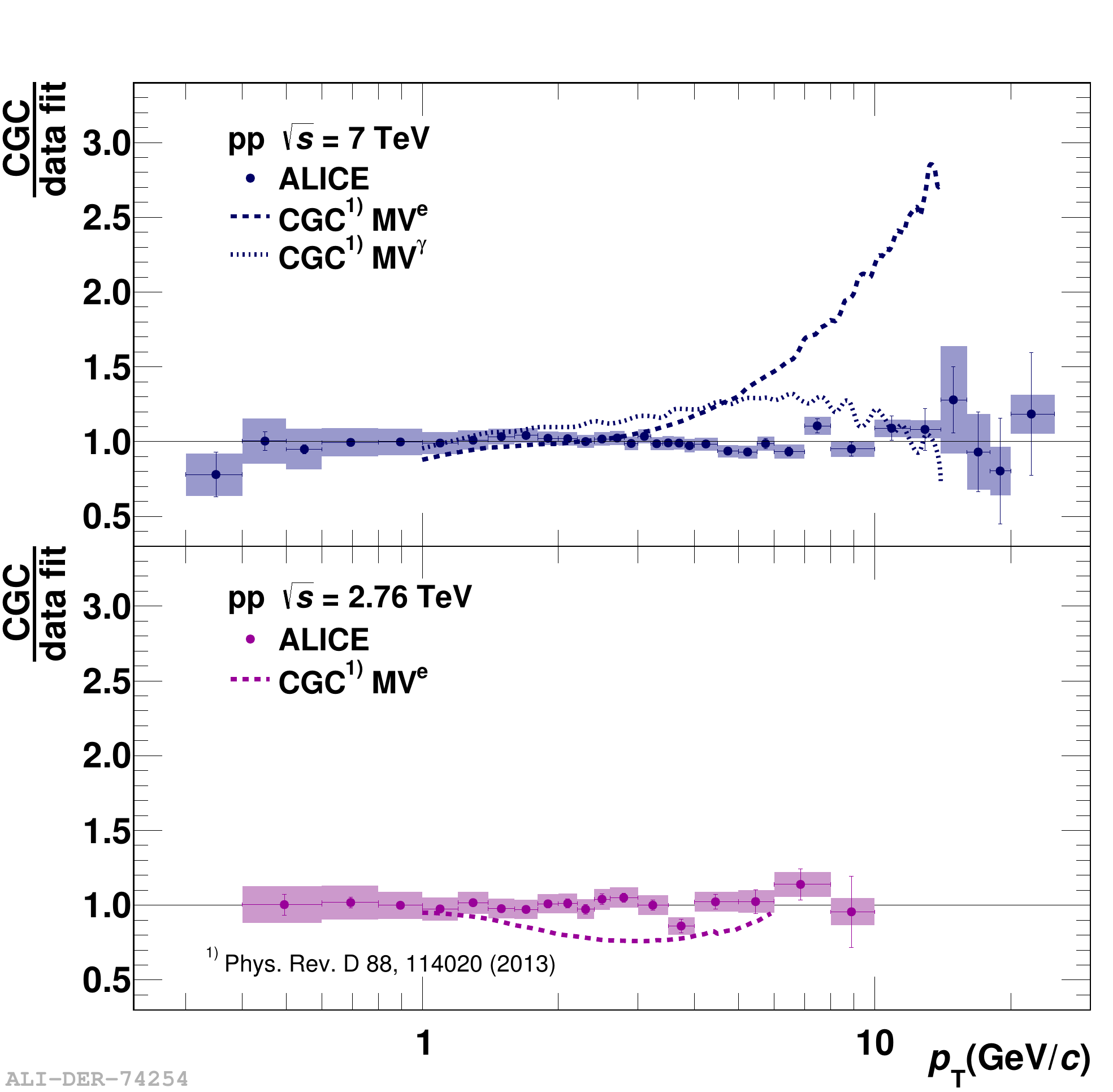}
\end{subfigure}\vspace{-0.4cm}
\caption{\footnotesize Left panel:invariant yields at ${\sqrt{s}=}$ 7~TeV and 2.76~TeV compared to CGC models. Right panel: ratio of the ALICE data fit to CGC predictions.}\label{fig:cgc}
\end{figure}

Further comparisons of the pp results above have been made to color glass condensate calculations (CGC) which include saturation of low momentum fraction ($x$) gluons. These comparisons in Fig.~\ref{fig:cgc} show that this CGC calculation is able to describe the data up to intermediate $\pt$ and only the MV$^{\gamma}$ parametrization (McLerran-Venugopalan model with anomalous dimension $\gamma$)~\cite{CGC} is able to describe up to the highest $\pt$.
\vspace{-0.1cm}

\vspace{-0.5cm}

%%%%%%%%%%%%%%
\section{$\piz$ in \pbpb and the nuclear modification factor $\raa$}
\begin{figure}[H]
\centering
\begin{subfigure}[b]{0.4\textwidth}
\includegraphics[width=4.0cm]{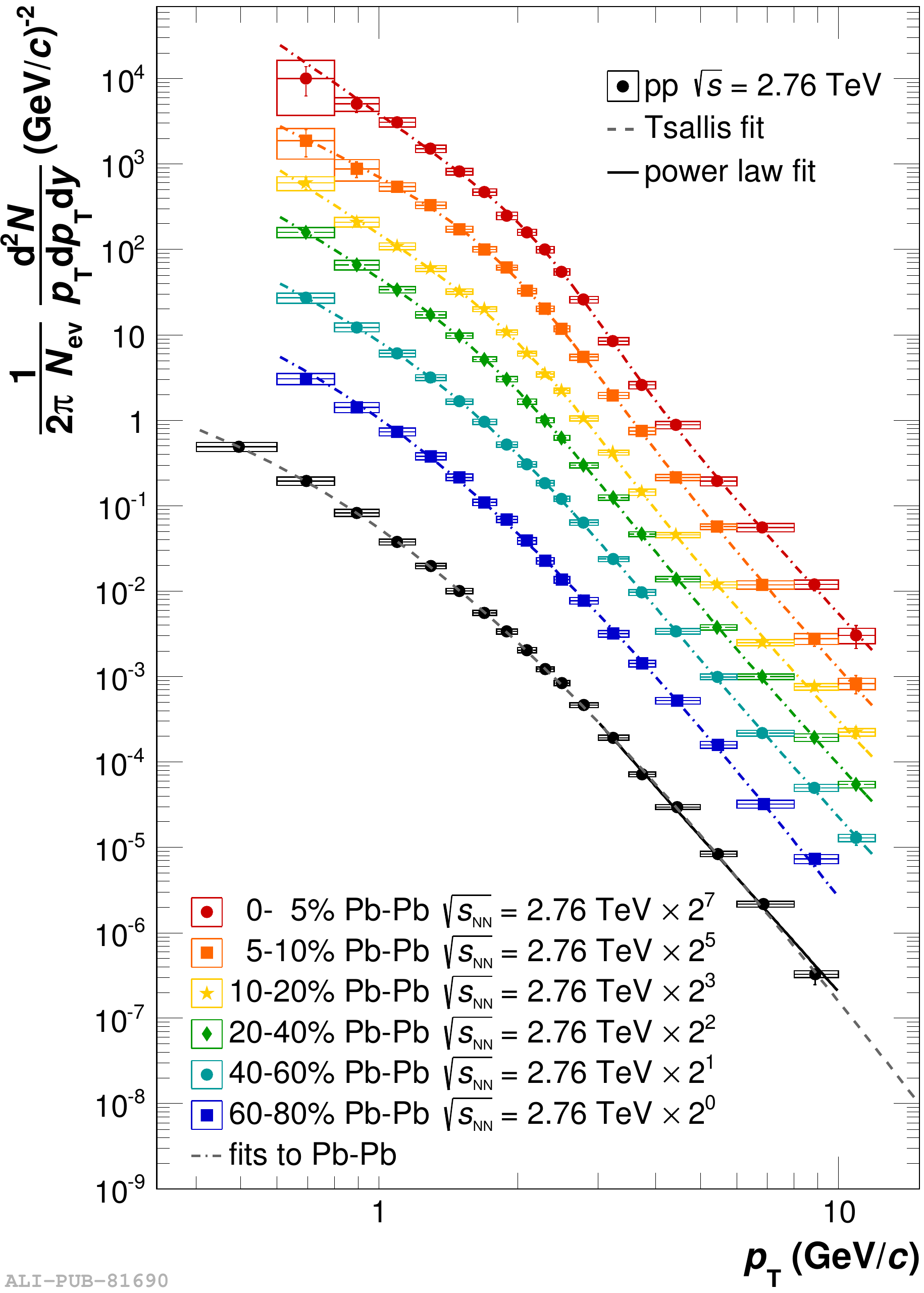}
\end{subfigure}
\begin{subfigure}[b]{0.4\textwidth}
\hspace{-1mm}\includegraphics[width=4.2cm]{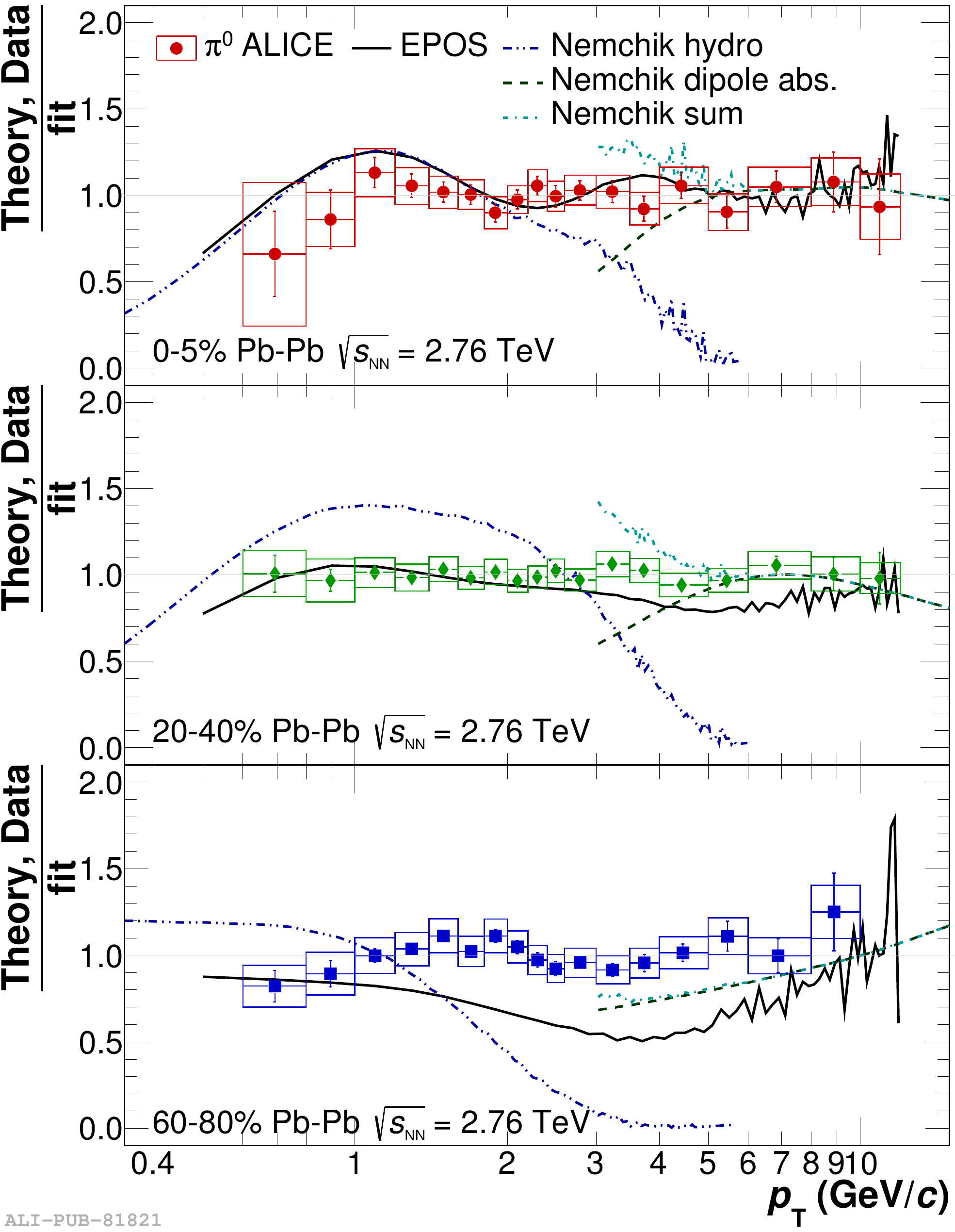}
\end{subfigure}
\caption{\footnotesize Left panel: $\piz$ invariant yields at ${\sqrt{s}=}$ 2.76~TeV in \pbpb collisions. Right panel: ratio of data and model/parametrization predictions.}\label{fig:pbpbepos}
\end{figure}

$\piz$ invariant yields have been measured by ALICE in six centrality classes: 0-5~$\%$, 5-10~$\%$, 10-20~$\%$, 20-40~$\%$, 40-60~$\%$ and 60-80~$\%$.
EPOS parametrizations which include hydrodynamic flow and energy loss of string segments~\cite{epos} seem to describe the data, in particular at the most central classes of ALICE measurements. Models which in addition of the hydrodynamic description also include color dipole absorption~\cite{nemchik}, fail to give an accurate description of the data trend as it can bee seen in Fig.~\ref{fig:pbpbepos} (right).

\begin{figure}[H]
\centering
\begin{subfigure}[b]{0.4\textwidth}
\includegraphics[width=5cm]{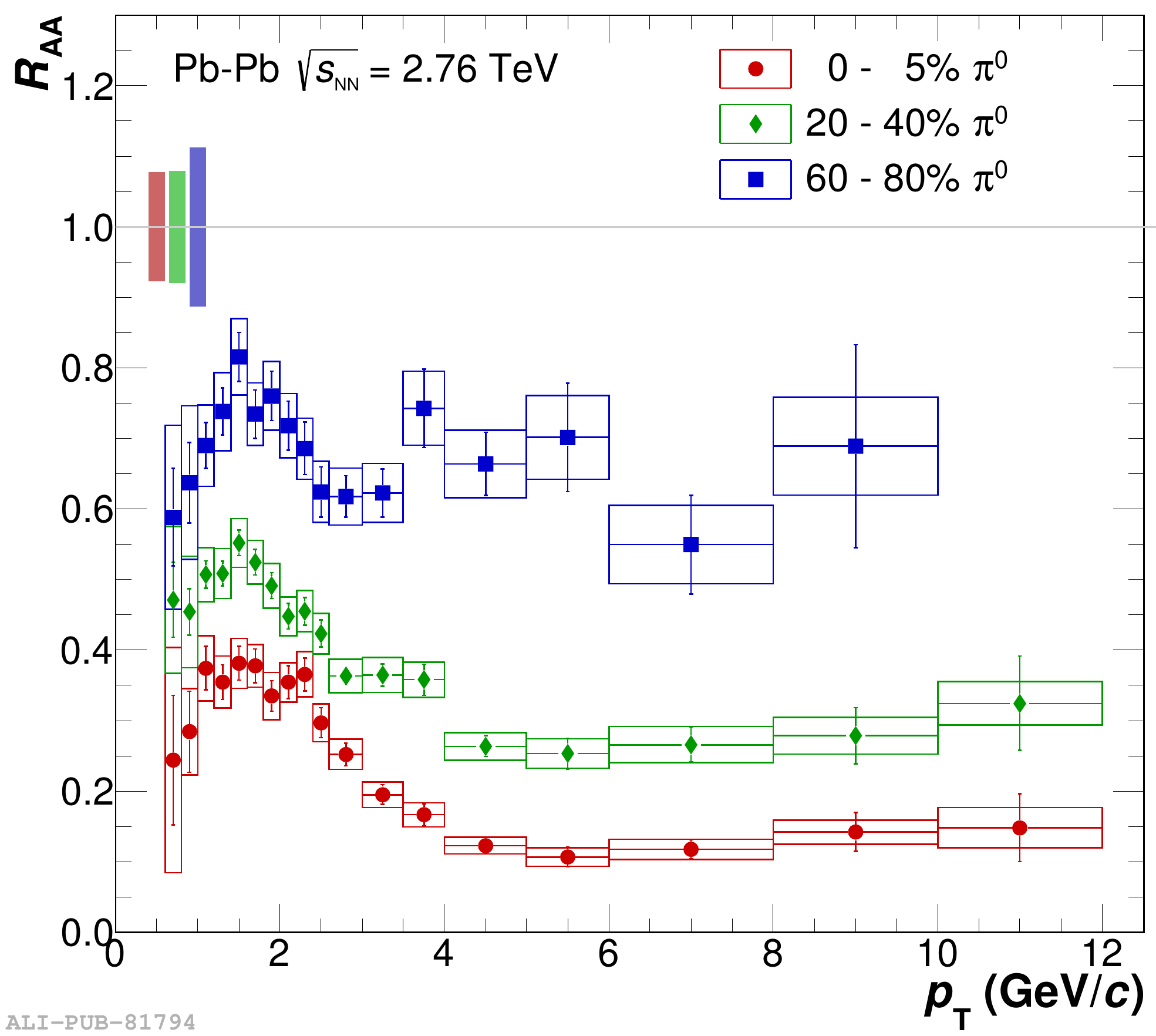}
\end{subfigure}
\begin{subfigure}[b]{0.4\textwidth}
\hspace{-1mm}\includegraphics[width=5cm]{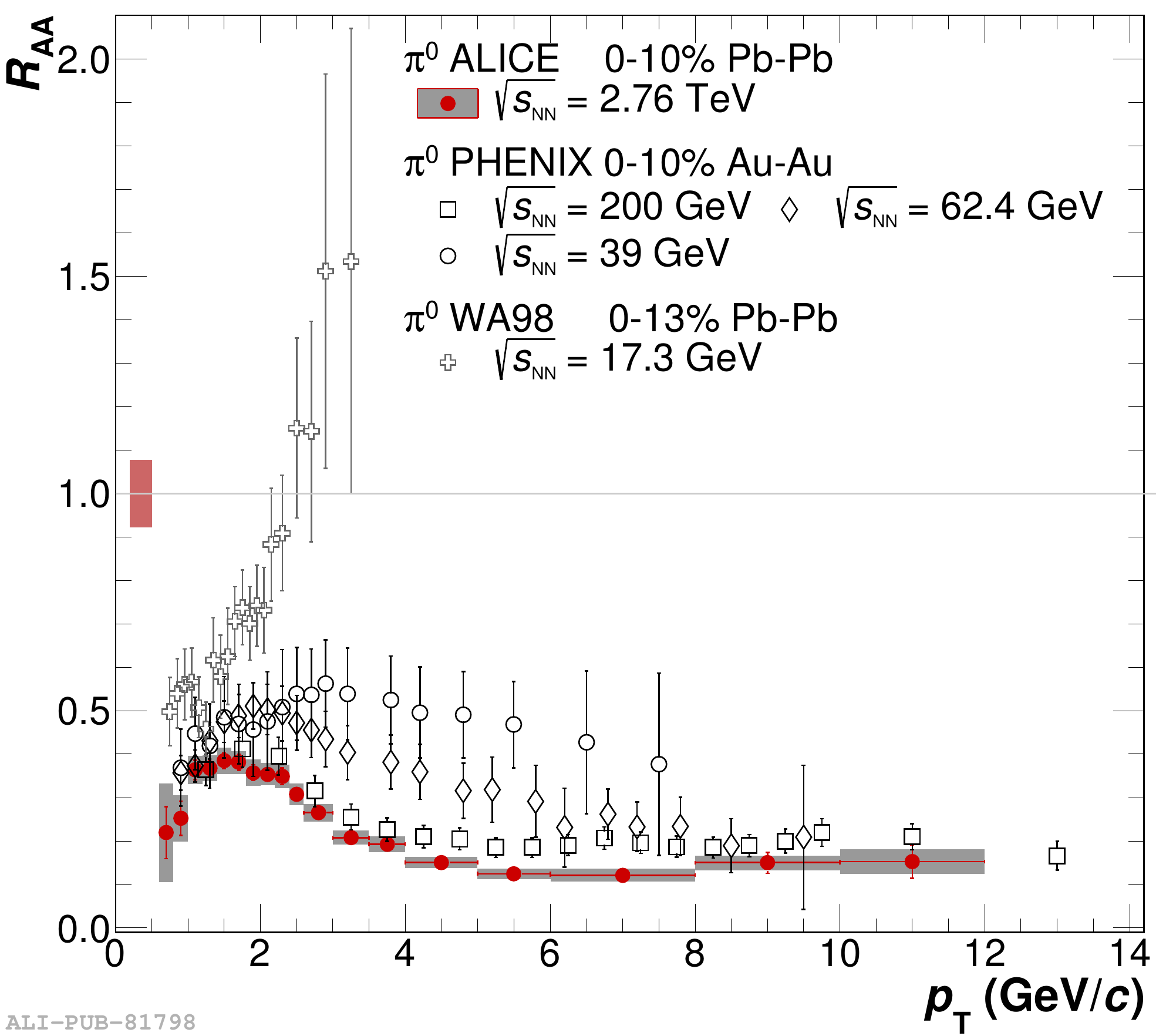}
\end{subfigure}
\caption{\footnotesize Left: $\piz$ invariant yields at ${\sqrt{s}=}$~2.76~TeV in \pbpb collisions. Right: ratio to model and parametrization comparisons (b).}\label{fig:raapbpb}
\end{figure}

A manner to quantify the nuclear effects in heavy ion (A-A) collisions is to measure the nuclear modification factor which is defined as: $R_{\aa}(\pt)=\frac{1}{N_{\rm coll}}\frac{dN_{\aa}/d\pt}{dN_{\rm pp}/d\pt}$. 
 $R_{\aa}$  compares the yields in Fig.~\ref{fig:pbpbepos} (left) to  production in scaled pp collisions from Fig.~\ref{fig:ppyields} at the same $\sqrt{s}$. This ratio is scaled by the number of binary nucleon-nucleon collisions (${N_{\rm coll}}$) which are in turn taken from Glauber Monte Carlo simulations~\cite{centrality}. ${R_{\aa}}$ is comprised of initial (Cronin, nuclear shadowing) and final state effects such as jet quenching. These efefcts would need the corresponding measurement in proton-heavy ion (p-A) collisions for proper disentanglement and interpretation. 
ALICE has measured ${\piz}$ $R_{\aa}$ in six centrality classes and has found 
a large ${\piz}$ suppression in central \pbpb collisions as it can be seen in Fig.~\ref{fig:raapbpb}. In addition ALICE data when compared to the world measurements at different $\sqrt{s}$~\cite{pbpbpublished, PHENIX1,PHENIX2, WA98}, indicates an energy dependence with the ALICE's highest $\sqrt{s}$ data points exhibiting the highest suppression (Fig.~\ref{fig:raapbpb} right).

\vspace{-0.5cm}
%%%%%%%%%%%%%%%%
\section{$\piz$-hadron correlations}
\begin{figure}[H]
 \begin{subfigure}[b]{0.26\textwidth}
\includegraphics[width=3.9cm]{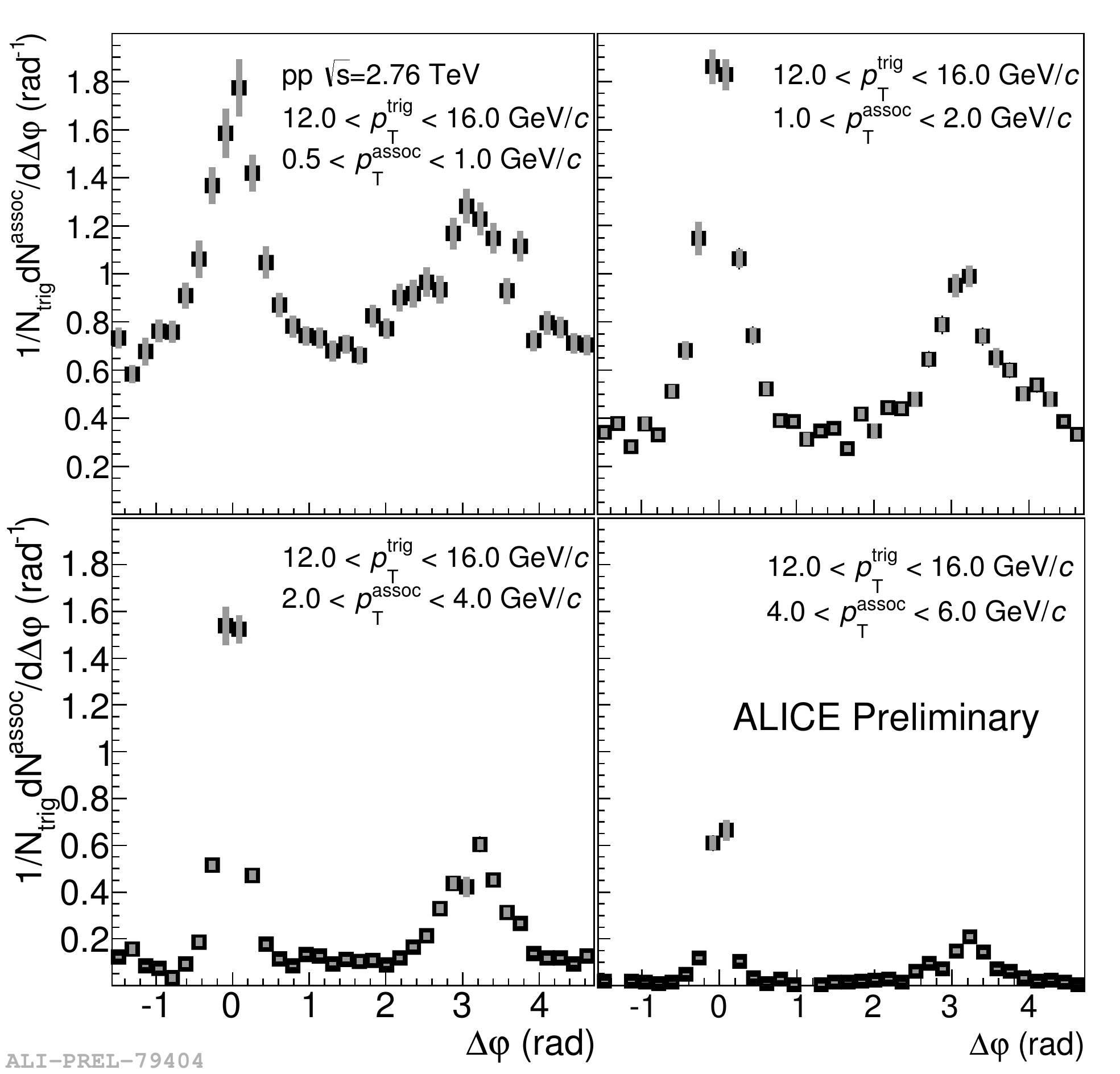}
\caption{\footnotesize pp collisions}\label{fig:a}
\end{subfigure}\begin{subfigure}[b]{0.26\textwidth}
\includegraphics[width=3.9cm]{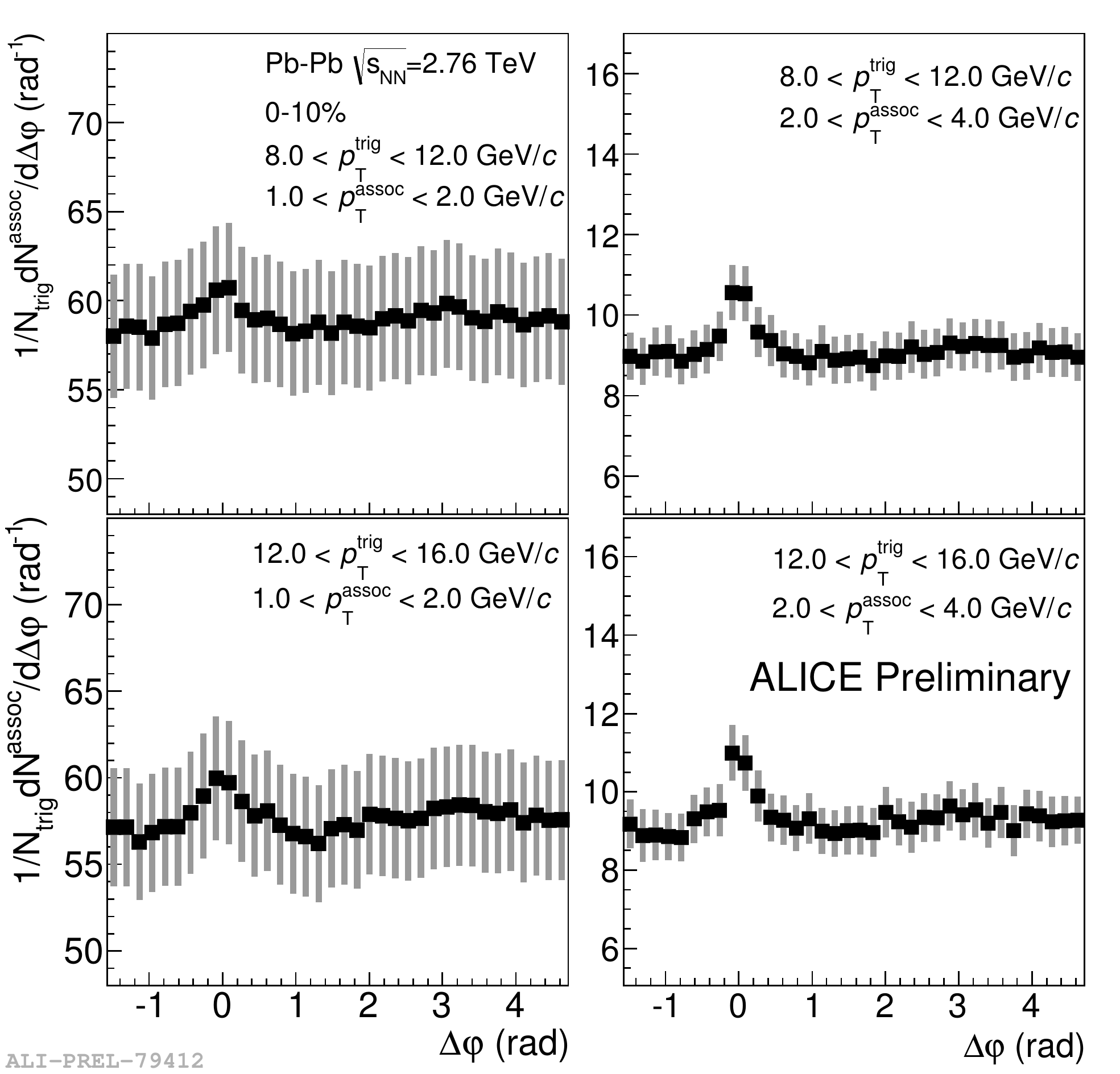}
\caption{\footnotesize \pbpb collisions}\label{fig:b}
\end{subfigure}\begin{subfigure}[b]{0.26\textwidth}
\includegraphics[width=3.9cm]{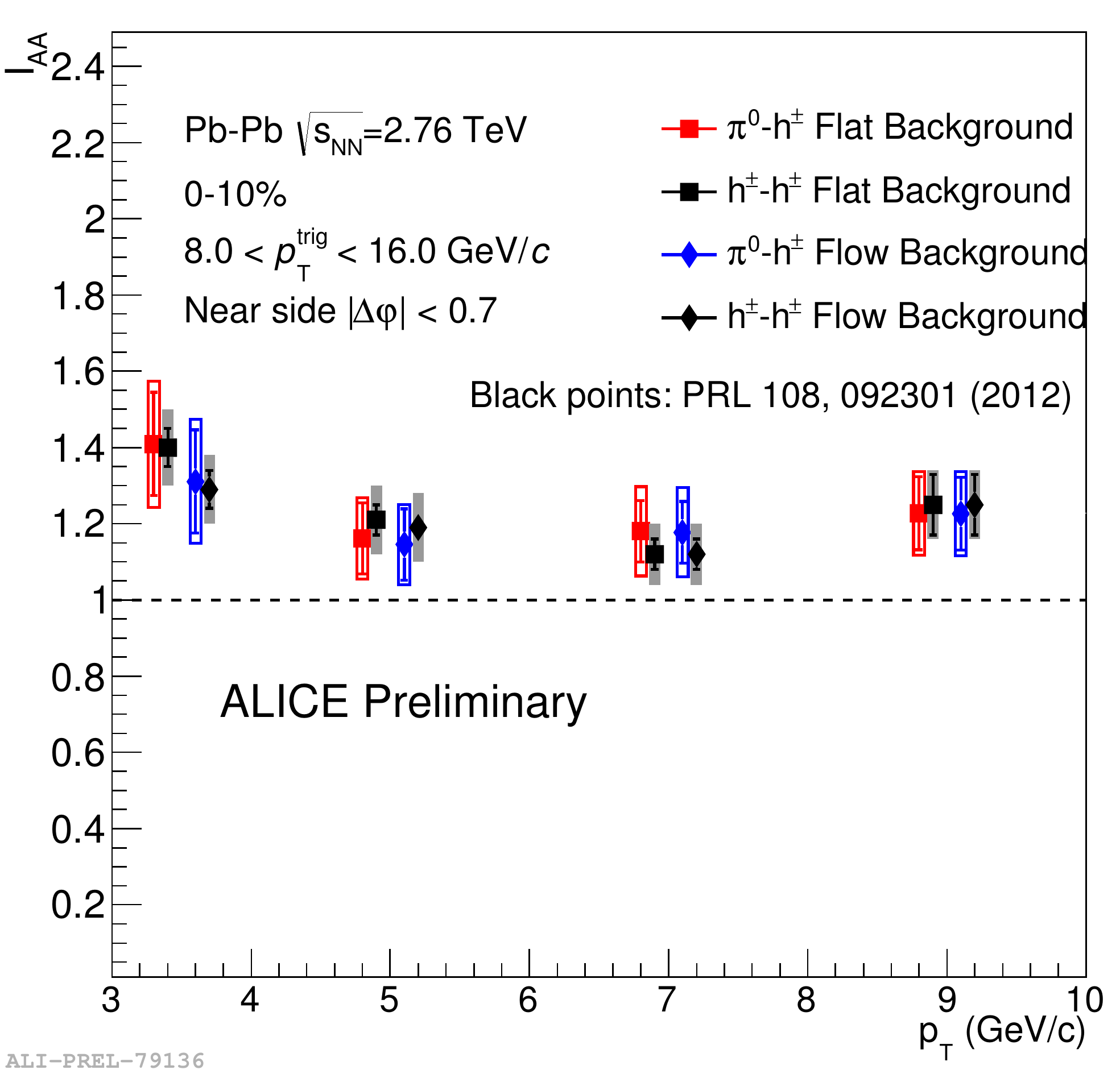}
\caption{\footnotesize Near side $|\Delta \phi|<0.7$ rad}
\end{subfigure}\begin{subfigure}[b]{0.26\textwidth}\label{fig:c}
\includegraphics[width=3.9cm]{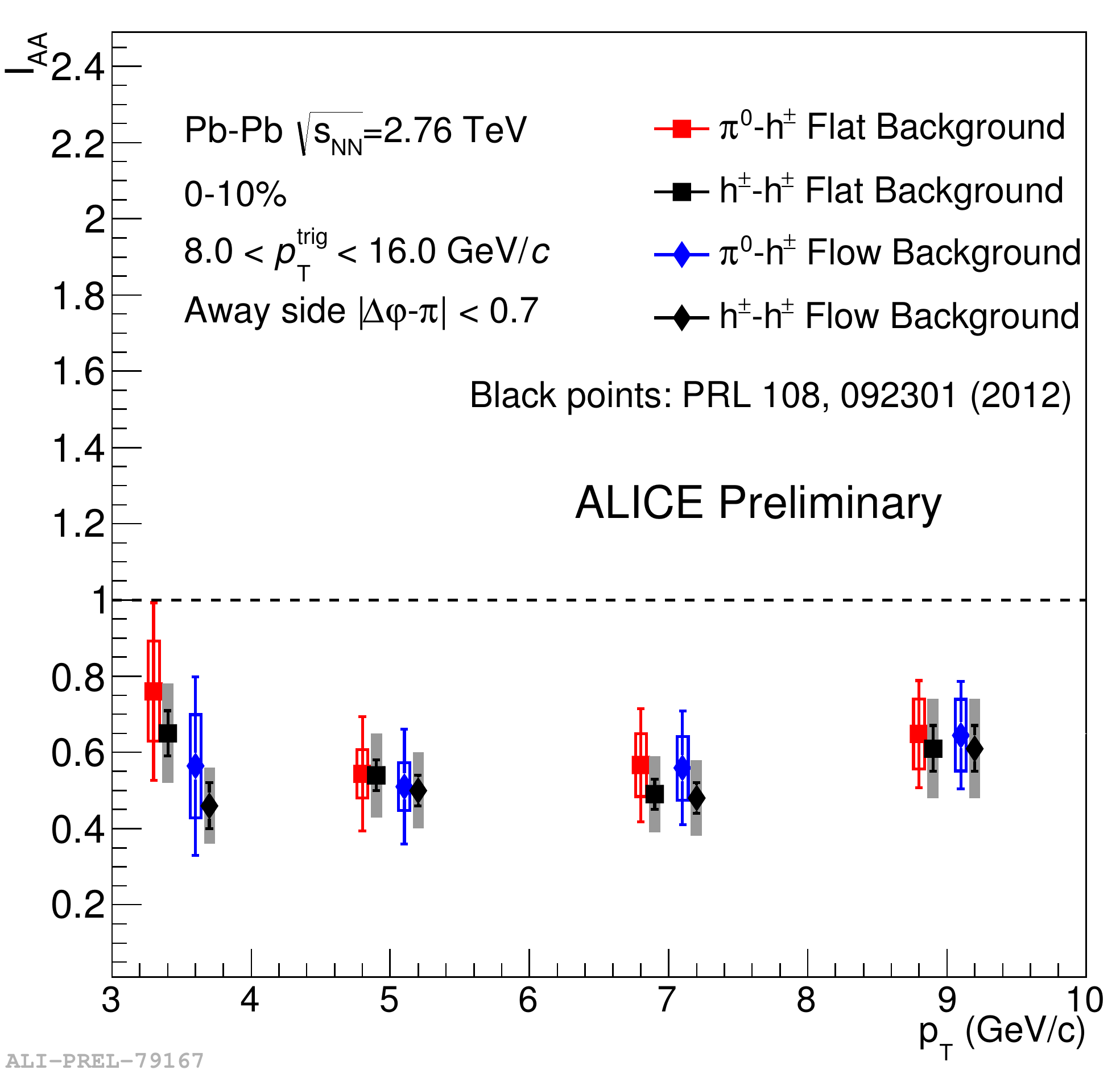}
\caption{\footnotesize Away side $|\Delta \phi-\pi|<0.7$ rad}\label{fig:c}
\end{subfigure}
\caption{\footnotesize (a) and (b)$: \piz$ -hadron azimuthal angle correlations. (c )and (d): $I_{\aa}(\pt^{\piz},\pt^{h^{\pm}})$.}\label{fig:pi0h}
\end{figure}

Correlations between high-$\pt$ photons or leading hadrons and charged hadrons are considered as a sensitive probe for studying medium-induced parton energy loss and jet modification in heavy-ion collisions. ALICE has measured the azimuthal angular correlations between neutral pions triggered by ALICE's EMCAL single shower trigger and charged hadrons detected in the central tracker.
Fig.~\ref{fig:pi0h} (a and b) indicates that a strong suppression of away-side correlation is observed in \pbpb central collisions with respect to pp collisions.
\vspace{-0.2cm}
\vspace{-0.2cm}
A similar observable where effects of the medium on the yield of particles in a jet can be studied is the per-trigger yield $I_{\aa}(\pt^{\piz},\pt^{h^{\pm}})=\frac{Y^{\pbpb}(\pt^{\piz},\pt^{h^{\pm}})}{Y^{\pp}(\pt^{\piz},\pt^{h^{\pm}})}$ . The latest $I_{\aa}$ results obtained by ALICE are in Fig.~\ref{fig:pi0h} (c and d). Results indicate an ${I_{\aa}}$ enhancement on the near side (Fig.~\ref{fig:pi0h} c), while a suppression is observed (Fig.~\ref{fig:pi0h} d) on the away side further supporting the picture of parton energy loss in the medium.
The fragmentation function modification measured via high-$\pt\,\piz$ is consistent with corresponding high-$\pt$ charged particle measurements performed by ALICE.

%%%%%%%
\vspace{-0.3cm}
\section{Direct photons: $R_{\gamma}$}
\begin{figure}[H]
\centering
\begin{subfigure}[b]{0.4\textwidth}
\includegraphics[width=5.4cm]{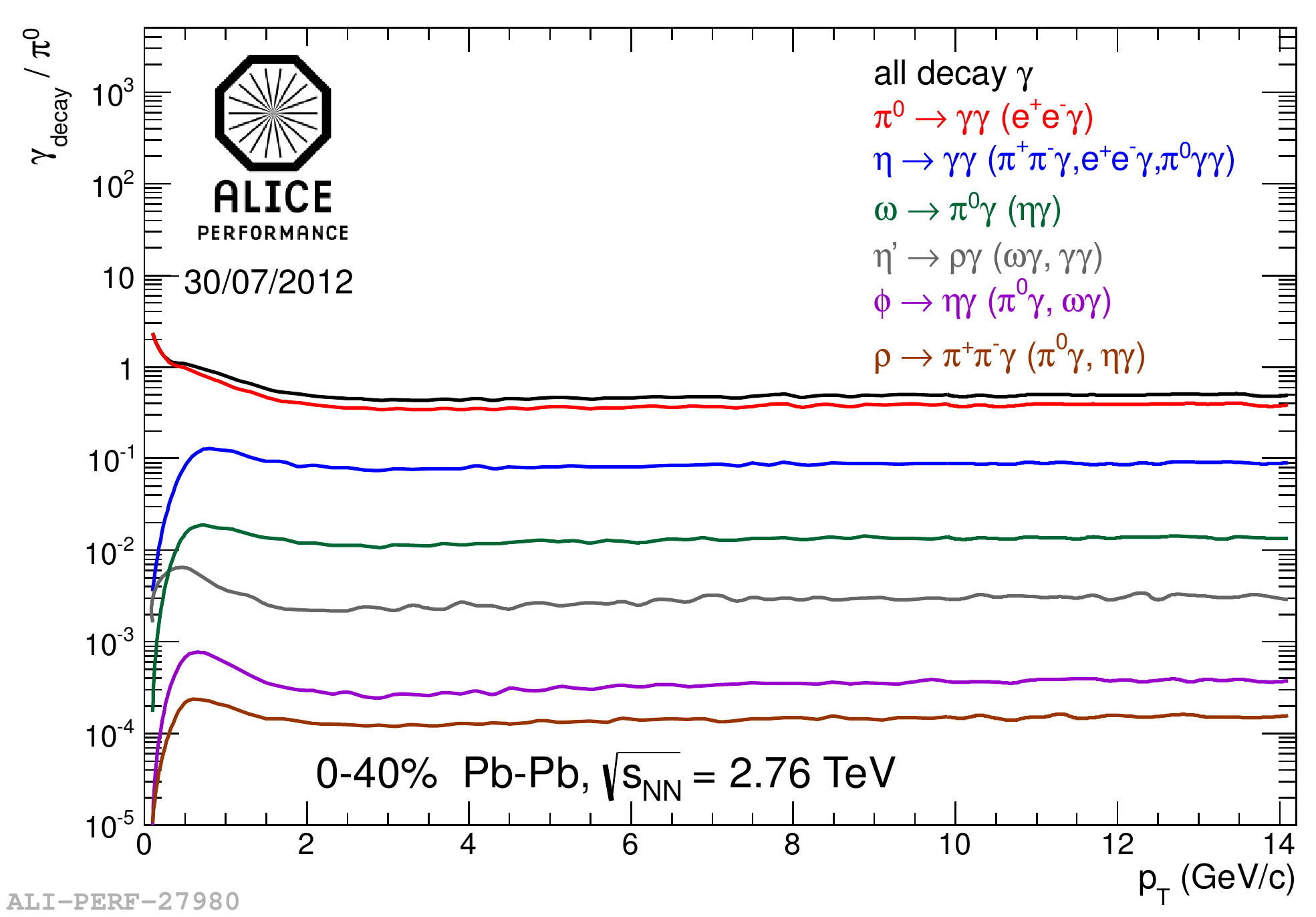}\caption{}
\end{subfigure}\begin{subfigure}[b]{0.4\textwidth}
\includegraphics[width=5.4cm]{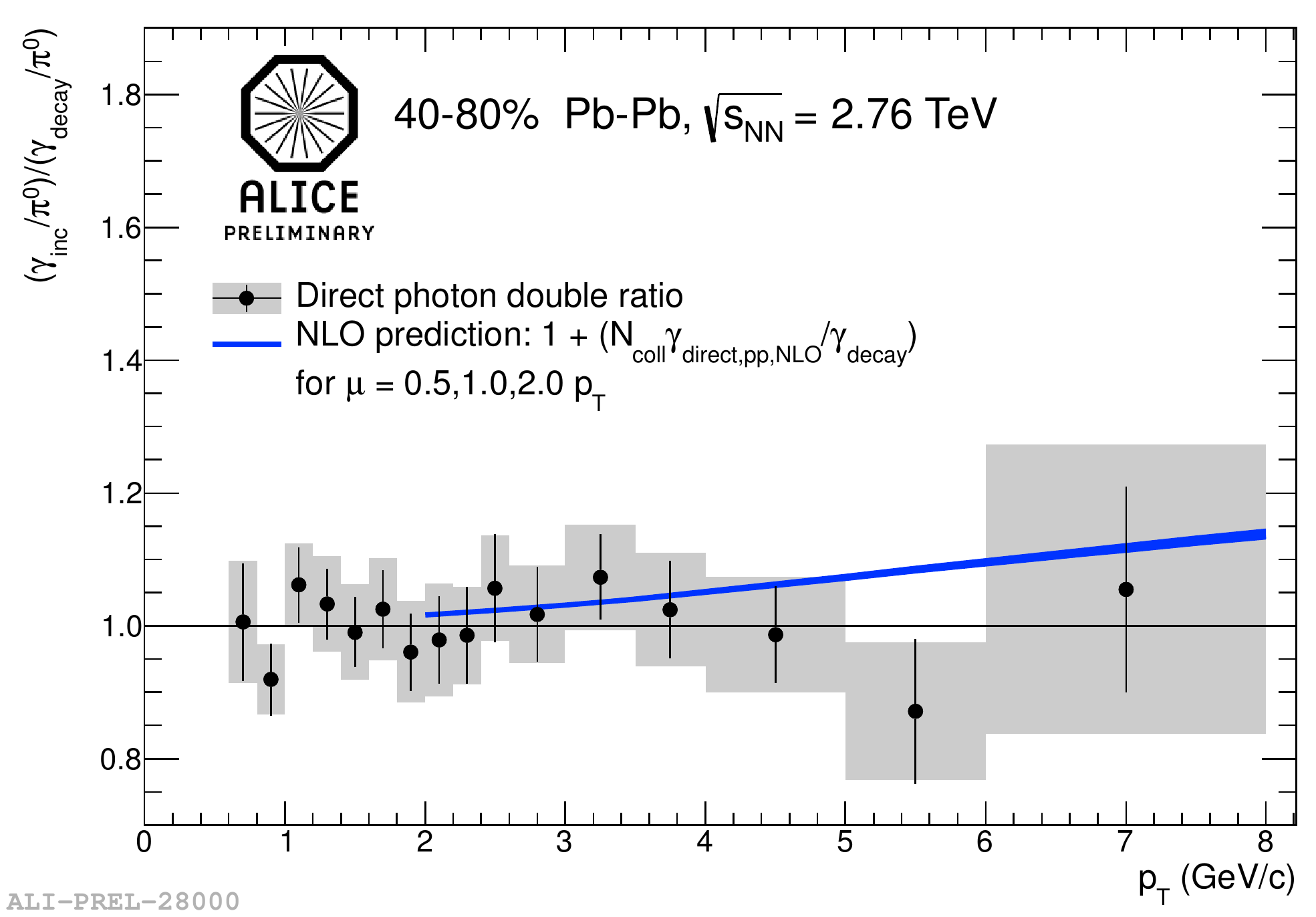}\caption{}
\end{subfigure}

\begin{subfigure}[b]{0.4\textwidth}
\includegraphics[width=5.4cm]{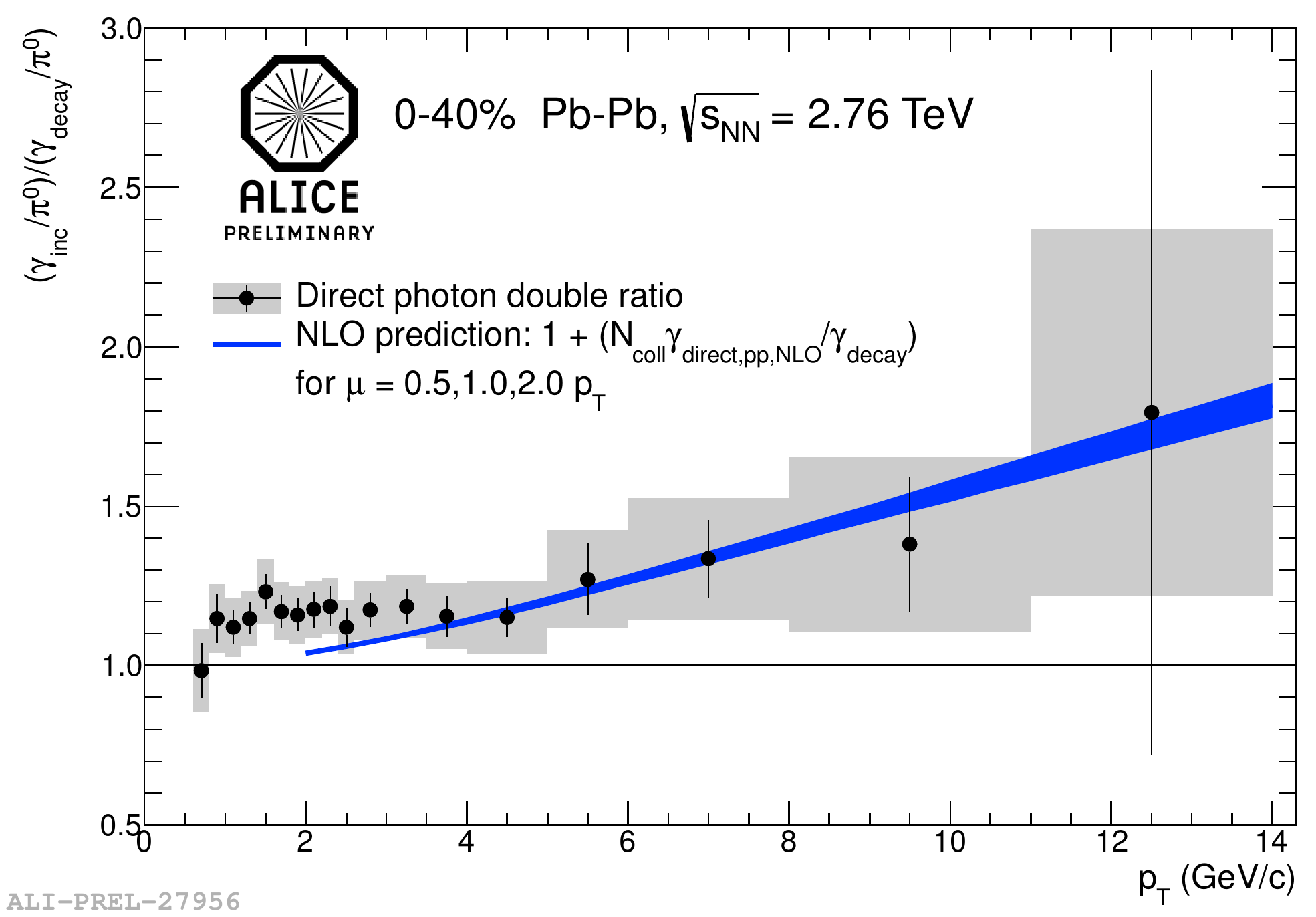}\caption{}
\end{subfigure}\begin{subfigure}[b]{0.4\textwidth}
\includegraphics[width=5.4cm]{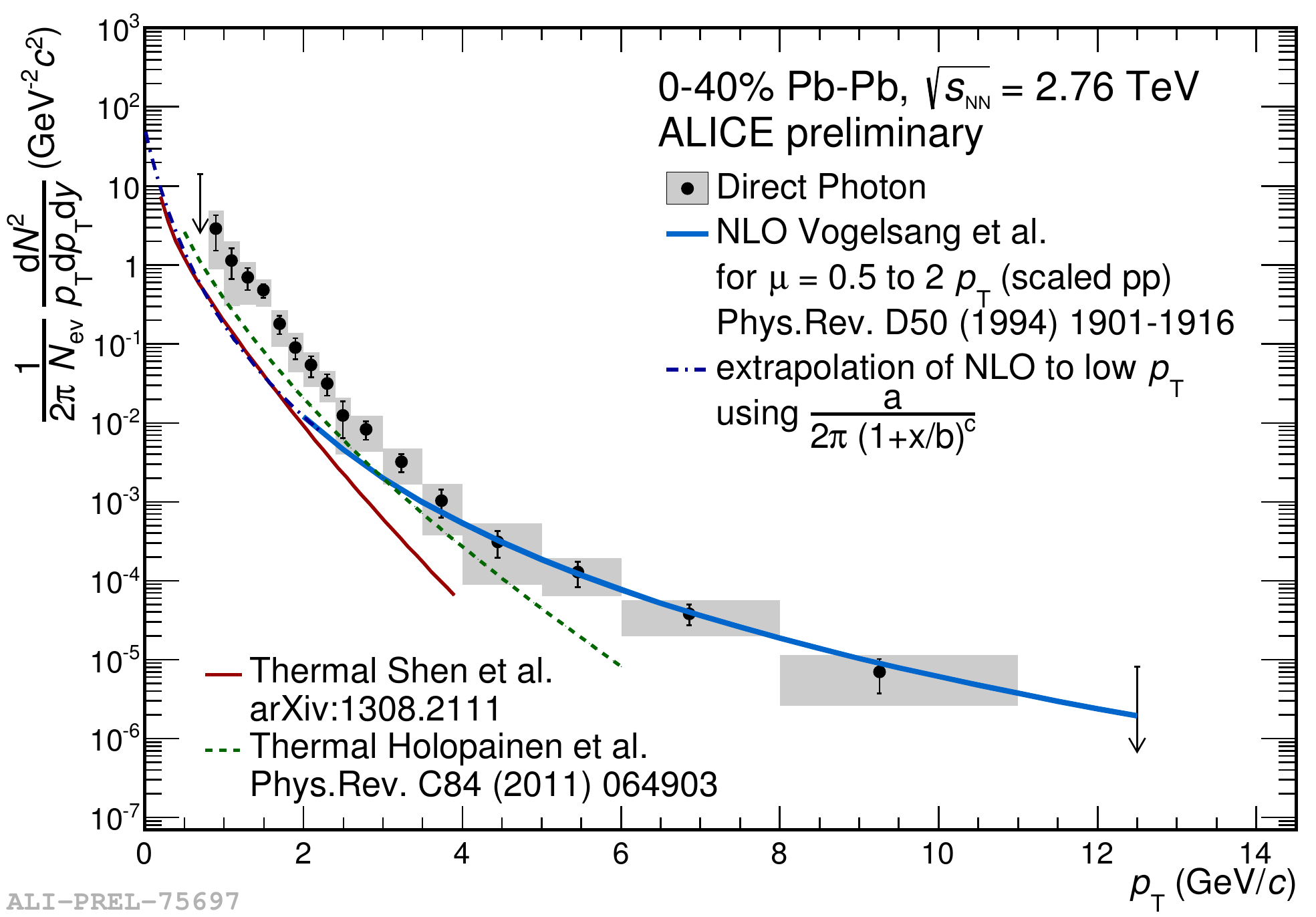}\caption{}
\end{subfigure}\vspace{-0.3cm}
\caption{\footnotesize (a): Particle decay photon contributions. (b): $R_{\gamma}$ in peripheral \pbpb collisions.  (c): $R_{\gamma}$ in central \pbpb collisions.  (d): Direct photon spectrum  derived from the double ratio by ${\gamma_{\rm direct}=(1-1/R_{\gamma})\gamma_{\rm inc}}$}\label{fig:directphoton}
\end{figure}

\vspace{-0.2cm}
The method for extracting a direct photon signal depends on the measurement of the inclusive photon yield (${\gamma_{\rm~inc}}$ ) via the reconstruction of their conversion products~\cite{directphoton}.
The direct-photon signal is extracted via a subtraction method defined as:
$\gamma_{\rm direct}=\gamma_{\rm~inc}-\gamma_{\rm decay}=(1-\frac{\gamma_{\rm decay}}{\gamma_{\rm inc}})\gamma_{\rm inc}
=(1-\frac{1}{R_{\gamma}})\gamma_{\rm inc}$.
Raw photon yields are corrected for purity, reconstruction efficiency and material photon conversion probability. The contribution to the spectrum due to decays: $\gamma_{\rm decay}$, is obtained from spectra parametrizations of mesons with photon decay branches (Fig.~\ref{fig:directphoton} a) using either measured data (when available) or $m_{\rm T}$ scaling (for the unmeasured sources).  The bulk of the contribution of these decays comes from $\piz$ decays ($\sim$ 80$\%$) and $\eta$ mesons $\sim$ 18$\%$. 

From the inclusive spectrum a double ratio is derived: ${R_{\gamma}=\frac{\gamma_{\rm inc}}{\piz}/\frac{\gamma_{\rm decay}}{\piz_{\rm param}}}$. The quantity ${ \frac{\gamma_{inc}}{\piz}}$ represents the inclusive photon spectrum per ${\piz}$ and ${\frac{\gamma_{decay}}{\piz_{param}}}$ are the decay photons per ${\piz}$.
The double ratio serves to minimize the systematic uncertainties and it quantifies the photon signal. A value greater than one indicates the observation of direct photons while a value around unity would indicate no direct photon signal. 
ALICE preliminary measurements~\cite{directphoton} of the double ratio in Fig.~\ref{fig:directphoton} b show that in peripheral collisions, as expected, no photon production excess is observed. In central collisions and at low ${\pt}$ (${\pt}<$4 GeV/c), an excess of $20\%\pm5\%^{\rm stat}\pm10\%^{\rm syst}$ has been measured as it can be seen in Fig.~\ref{fig:directphoton} c. Comparisons of theoretical calculations have been made in Fig.~\ref{fig:directphoton} d however the current uncertainties in the data do not allow to discriminate between predictions beyond 2${\sigma}$.
\vspace{-0.2cm}

\vspace{-0.9cm}

%%%%%%%%%%%%%
\section{Direct ${\gamma}$ elliptic flow ${v_{2}}$ and ${v_{3}}$}
\vspace{-0.3cm}

The elliptic flow ${v_{2}}$ characterizes azimuthal anisotropies of photon emission with respect to the reaction plane and its obtained via the Fourier transform of the azimuthal distribution of particles $\frac{dN}{d\phi}=\frac{1}{2\pi} (1+2\sum_{n\geq1} v_{n}\cos(n(\phi-\psi^{RP}_{n}))$.
A ${v_{2}}$ signal at low ${\pt}$ would point towards collective behaviour. 
%A flow signal at high ${\pt}$  would perhaps indicate path length and in-medium parton energy loss. 
An observation of a thermal photon ${v_{2}}$ signal would additionally constrain the onset of direct photon production.
Small flow would be associated with early production while a large, hadron-like flow would point towards late production.

\begin{figure}[H]
\centering
\begin{subfigure}[b]{0.4\textwidth}
\includegraphics[width=4.5cm]{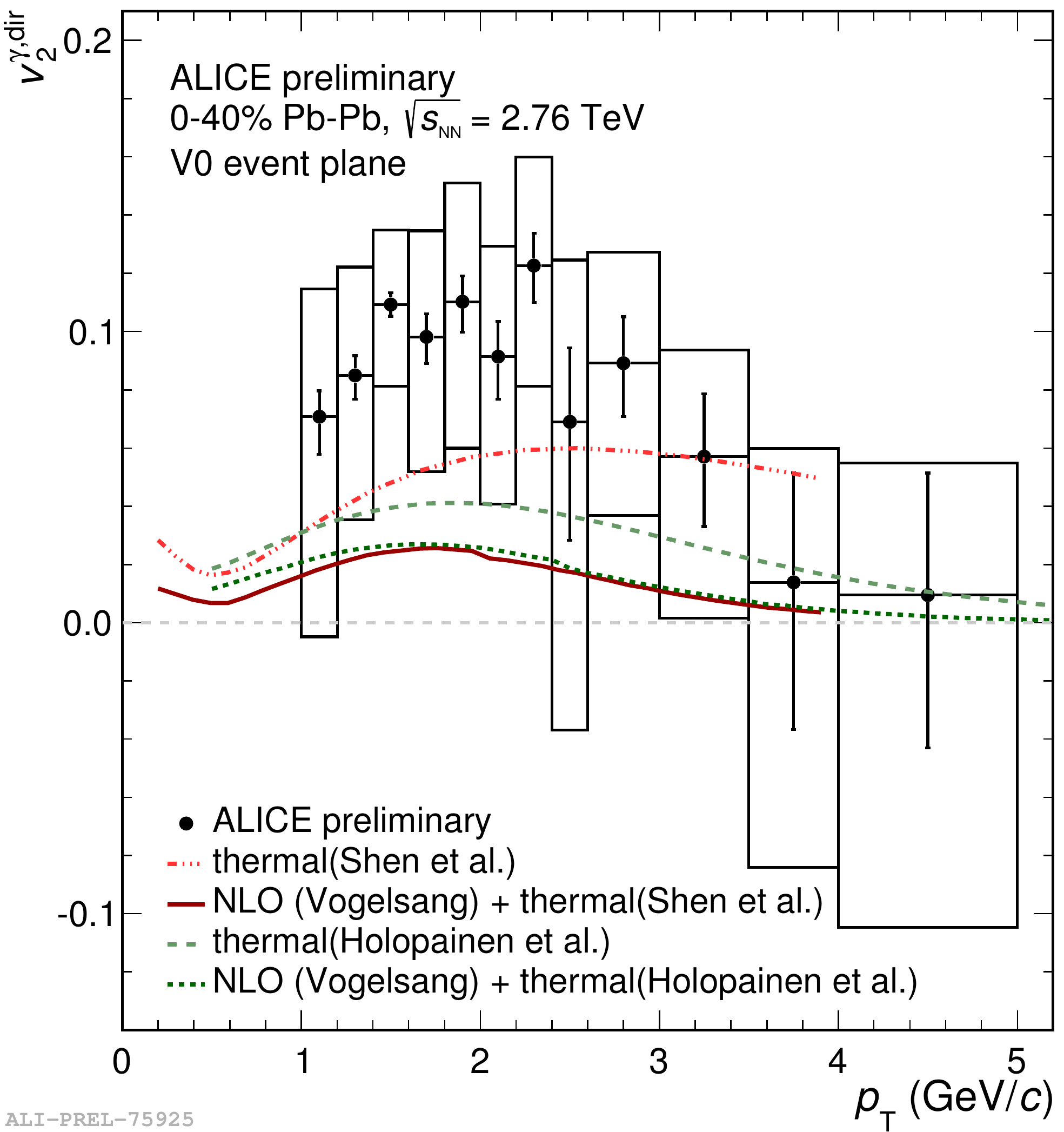}
\end{subfigure}\begin{subfigure}[b]{0.3\textwidth}
\includegraphics[width=4.5cm]{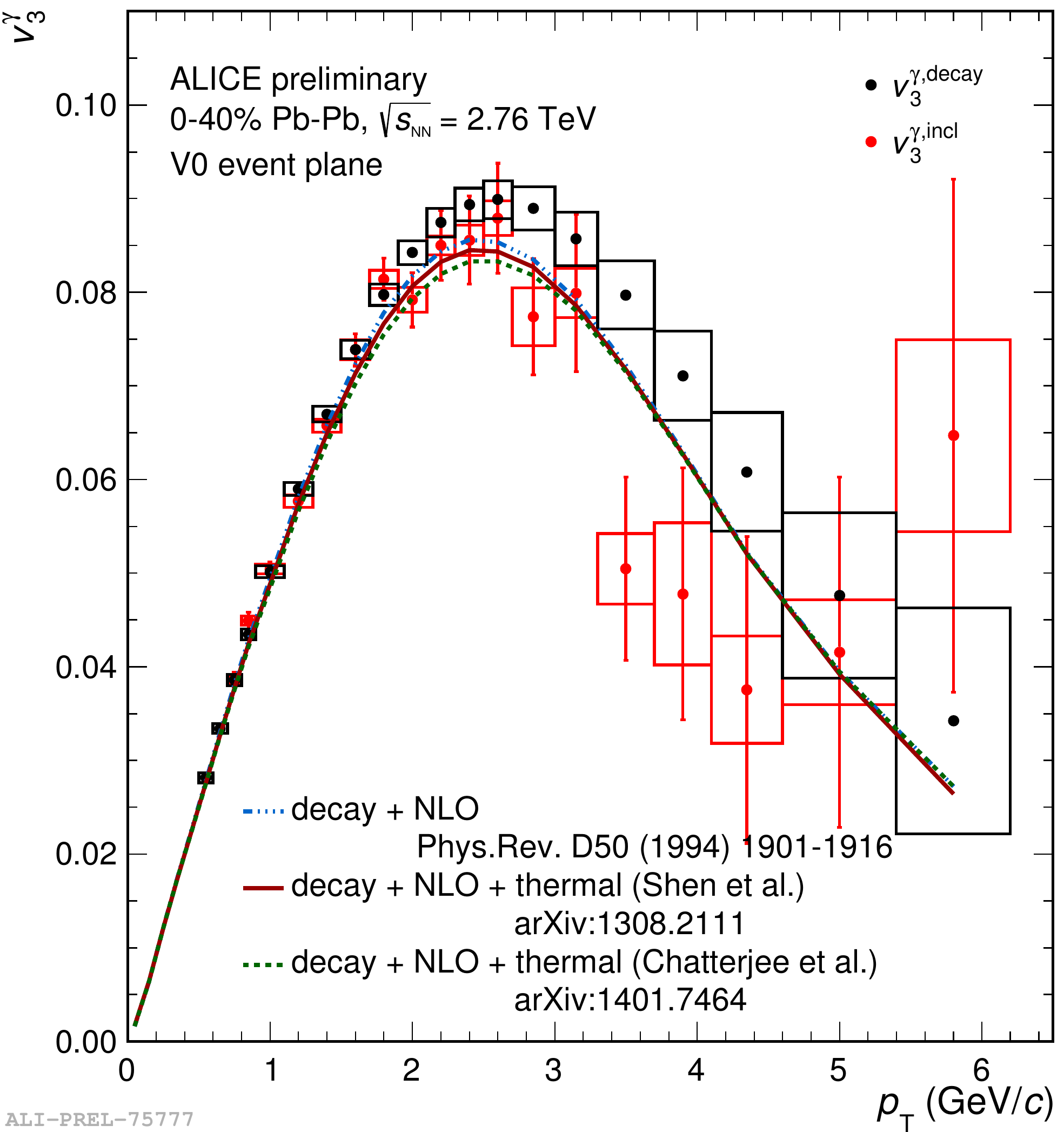}
\end{subfigure}\caption{\footnotesize Left panel: $v_{2}^{\rm inclusive~ \gamma}$. Right panel:  $v_{3}^{\rm inclusive~ \gamma}$}\label{fig:directphotonv23}
\end{figure}

Direct photon ${v_{2}}$ is calculated via: 
${v_{2}^{\rm direct\gamma}=\frac{R_{\gamma}v_{2}^{\rm inc\gamma}-v_{2}^{\rm  decay\gamma}}{R_{\gamma}-1}}$,  where ${R_{\gamma}\,v_{2}^{\rm inc\,\gamma}}$ is the weighted inclusive photon spectrum ${v_{2}}$ due to extra photons compared to background. ${v_{2}^{\rm decay\gamma}}$ is the calculated decay photon ${v_{2}}$ from the cocktail calculation as it was the case for the direct photon invariant yields (Fig.~\ref{fig:directphoton} a). Since both measurements of invariant yields and ${v_{2}}$ depend on ${R_{\gamma}}$, the significance of the excess seen in Fig.~\ref{fig:directphotonv23} (left) depends on the critical assessment of systematics and their propagation, both which are still under investigation.

$v_{3}$ is the next term in the Fourier transform of the azimuthal distribution of particles.
ALICE has measured the inclusive photon ${v_{3}}$ which is in turn a first time measurement at the LHC. The data shown in  Fig.~\ref{fig:directphotonv23} (right)  indicates that for $\pt > 3$~GeV/c, direct photon ${v_{3}}$ contribution consistent with ${v^{\rm direct}_{3}}$ $<$ ${v^{\rm decay}_{3}}$ as expected for prompt photons, albeit with large statistical uncertainties. The same data below 3 GeV/c is mostly consistent within uncertainties. 
\vspace{-1.0cm}
\section{Summary and conclusions}\vspace{-0.2cm}
${\piz}$ invariant yields have been measured by ALICE in pp at three center of mass energies and in 6 centrality classes in {\pbpb} collisions at $\sqrt{s_{NN}}=$~2.76~TeV.
The results show that NLO pQCD calculations do not describe well ${\piz}$ production in \pp collisions at higher center of mass energies (${\sqrt{s}} =$ 2.76 and 7~TeV).
A suppression and a $\sqrt{s}$ dependence is observed for the measured ${\piz}$'s ${R_{\aa}}$.
The observed suppression is stronger at the LHC where the collision energy is higher than what had been previously studied. 
Theoretical models concerning ${\piz}$ production in \pbpb collisions only partially describe ALICE data.
${\piz}$-hadron correlations have been measured in pp and \pbpb collisions. The results indicate an away-side suppression in central collisions pointing towards the observation of parton energy loss.
Direct photon ${R_{\gamma}}$ and invariant yields have been measured in two centrality classes.
%with an exponential slope of T $=$ 304 $\pm$ 51 stat$+$syst MeV 
The interpretation of these results are still awaiting a more detailed treatment of the current systematics. Direct photon ${v_{2}}$ and a first measurement of inclusive photon ${v_{3}}$ at the LHC has been measured in {\pbpb} collisions in the 0-40$\%$ centrality class. Due to the current statistical limitations, which are expected to improve with the upcoming re-start of LHC operations, no conclusions can be made on $v_{2}$ and $v_{3}$ . 
\vspace{-0.5cm}
 
%%%%%%%%%%%%%%%%%%%%%%%%%%%%%%%%%%%%%%%%%%%%%%%%%%%%%%%%%%%%%%%%%%%%%%%%%
%%
%%   use this format to include an .eps figure into your paper
% %%
% \begin{figure}[htb]
% \centering
% \includegraphics[height=2in]{head_lhcp2014.jpg}
% \caption{ Place the caption here}
% \label{fig:figure1}
% \end{figure}
%%%%%%%%%%%%%%%%%%%%%%%%%%%%%%%%%%%%%%%%%%%%%%%%%%%%%%%%%%%%%%%%%%%%%%%%%%%

%%%%%%%%%%%%%%%%%%%%%%%%%%%%%%%%%%%%%%%%%%%%%%%%%%%%%%%%%%%%%%%%%%%%%%%%%
%%
%%   use this format to include a LaTeX table  into your paper
%%
% \begin{table}[t]
% \begin{center}
% \begin{tabular}{l|ccc}  
% Patient &  Initial level($\mu$g/cc) &  w. Magnet &  
% w. Magnet and Sound \\ \hline
%  Guglielmo B.  &   0.12     &     0.10      &     0.001  \\
%  Ferrando di N. &  0.15     &     0.11      &  $< 0.0005$ \\ \hline
% \end{tabular}
% \caption{ place the caption here }
% \label{tab:table1}
% \end{center}
% \end{table}
%%%%%%%%%%%%%%%%%%%%%%%%%%%%%%%%%%%%%%%%%%%%%%%%%%%%%%%%%%%%%%%%%%%%%%%%%%%

\vspace{-1.4cm}
 

\begin{thebibliography}{99}
\footnotesize
%%
%%  bibliographic items can be constructed using the LaTeX format in SPIRES:
%%    see    http://www.slac.stanford.edu/spires/hep/latex.html
%%  SPIRES will also supply the CITATION line information; please include it.
%%

\bibitem{QCD}
Greiner, W {\it et al.}, 
Quantum Chromodynamics (Book)
Harry Deutsch, Thun, 1984, 1989.

\bibitem{emcal}
 B. Alessandro{\it et al.} [ALICE Collaboration]
ALICE Physics Performance Report 
J. Phys. G: Nucl. Phys. 2 (2006) 

\bibitem{QCD2}
Sarkar, S. {\it et al.}, 
The Physics of the Quark Gluon Plasma (Book)
Introductory Lectures Series: Lecture Notes in Physics, Vol. 785
Springer, 2010.


\bibitem{thermalphoton} 
S.~Turbide,{\it et al.}
Hadronic production of thermal photons
Phys. Rev.C69, 014903 (2004) 


\bibitem{phos}
 G. Dellacasa {\it et al.}, 
Photon Spectrometer PHOS, Technical Design Report. 
CERN/LHCC 99-4, 5 March 1999.



\bibitem{pppublished}
B.~Abelev {\it et al.}  [ALICE Collaboration],
  Phys.\ Lett.\ B {\bf 717}, 162 (2012)

\bibitem{pbpbpublished}
 B.~B.~Abelev {\it et al.}  [ALICE Collaboration],
 arXiv:1405.3794 [nucl-ex].
\bibitem{centrality}
 B.~Abelev {\it et al.}  [ALICE Collaboration],
 Phys.\ Rev.\ C {\bf 88}, no. 4, 044909 (2013)
\bibitem{CGC}
T. Lappi, H.Mäntysaari, Phys. Rev. D88 (2013) 114020

\bibitem{epos}
 K.~Werner, I.~Karpenko, M.~Bleicher, T.~Pierog and S.~Porteboeuf-Houssais,
Phys.\ Rev.\ C {\bf 85}, 064907 (2012)

\bibitem{nemchik}
 Nemchik, 
Phys.\ Rev.\ C {\bf 86}, 054904, 2012

\bibitem{PHENIX0}
 A.~Adare {\it et al.}  [PHENIX Collaboration],
 Phys.\ Rev.\ D {\bf 83}, 052004 (2011)
\bibitem{PHENIX1}
  A.~Adare {\it et al.}  [PHENIX Collaboration],
Phys.\ Rev.\ Lett.\  {\bf 109}, 152301 (2012)

\bibitem{PHENIX2}
A.~Adare {\it et al.}  [PHENIX Collaboration],
Phys.\ Rev.\ Lett.\  {\bf 101}, 232301 (2008)

\bibitem{WA98}
 M.~M.~Aggarwal {\it et al.}  [WA98 Collaboration],
Phys.\ Rev.\ Lett.\  {\bf 100}, 242301 (2008)

\bibitem{directphoton}
 M.~Wilde [on behalf of the ALICE Collaboration],
 Nucl.\ Phys.\ A {\bf 904-905}, 573c (2013)

\end{thebibliography}
\end{document}